\documentclass[eqsecnum,10pt,a4paper]{article}
\usepackage{amsmath}
\numberwithin{equation}{section}

\begin{document}

\title{A Simple Quantum-Mechanical Model of Spacetime II: Thermodynamics of Spacetime}

\author{Jarmo M\"akel\"a\footnote{Vaasa University of Applied Sciences,
Wolffintie 30, 65200 Vaasa, Finland, email: jarmo.makela@puv.fi}}

\maketitle

\begin{abstract}

  In this second part of our series of two papers, where spacetime is modeled by a graph,
where Planck-size quantum black holes lie on the vertices, we consider the thermodynamics 
of spacetime. We formulate an equation which tells in which way an accelerating, spacelike 
two-surface of spacetime interacts with the thermal radiation flowing through that surface.
In the low temperature limit, where most quantum black holes constituting spacetime are 
assumed to lie in the ground state, our equation implies, among other things, the Hawking 
and the Unruh effects, as well as Einstein's field equation with a vanishing cosmological 
constant for general matter fields. We also consider the high temperature limit, where the
microscopic black holes are assumed to lie in highly excited sates. In this limit our 
model implies, among other things, that black hole entropy depends logarithmically on its
area, instead of being proportional to the area.  

\end{abstract}



\maketitle

\section{Introduction}

 In the first part \cite{yksi} of our series of two papers we constructed a microscopic model of 
spacetime, where microscopic quantum black holes were used as the fundamental building 
blocks of space and time. Spacetime was assumed to be a graph, where black holes lie 
on the vertices. The only physical degree of freedom associated with a microscopic 
quantum black hole acting as a fundamental constituent of spacetime was assumed to be
its horizon area, and our idea was to reduce all properties of spacetime back to the
quantum-mechanical eigenvalues of the horizon areas of the holes. The horizon area
eigenvalues were assumed to be of the form
\begin{equation}
A_n = (n+\frac{1}{2})32\pi\ell_{Pl}^2,
\end{equation}
where $n=0,1,2,...$, and 
$\ell_{Pl}:=\sqrt{\frac{\hbar G}{c^3}}\approx 1.6\times 10^{-35}m$ is the Planck 
length. We focussed our attention to the objects which we called as two-dimensional 
subgraphs, and which may be viewed as discrete analogues of two-surfaces of
spacetime. Assuming that every microscopic quantum black hole lying on a two-dimensional
subgraph contributes to that graph an area, which is proportional to the
quantum number $n$ we found that in the low temperature limit, where most black
holes are assumed to be in the ground state, where $n=0$, the two-dimensional subgraph
possesses an entropy
\begin{equation}
S = \frac{\ln 2}{\alpha}A,
\end{equation}
where $A$ is the total area of the two-dimensional subgraph under consideration, and 
$\alpha$ is a numerical constant of order unity. In other words, we found that in the 
low temperature limit the entropy of a two-dimensional subgraph is proportional to its area. 
When written in the SI units, Eq.(1.2) takes the form:
\begin{equation}
S = \frac{\ln 2}{\alpha}\frac{k_B c^3}{\hbar G}A.
\end{equation}

  Eq.(1.3) was one of the most important results of our first paper. Its importance lies
in its close relationship with the Bekenstein-Hawking entropy law, which  states that black
hole possesses an entropy, which is proportional to its event horizon area.
\cite{kaksi, kolme} Eq.(1.3) gives 
a rise to the hopes that our quantum mechanical model of spacetime might not only be 
capable to provide a microscopic explanation to the Bekenstein-Hawking entropy law, but
it could also be used to predict new, unexpected properties of gravitation and spacetime.

  As such as it stands, however, Eq.(1.3) provokes several questions: What is the precise
value of $\alpha$? How does spacetime behave in the high temperature limit, where the 
microscopic quantum black holes, instead of lying close to the ground state, are assumed
to lie in highly excited states? Finally, we have the most important question of all: Is
it possible to obtain, at least at an appropriate limit, Einstein's field equation, and
thereby the classical general relativity with all of its consequences from our microscopic 
model of spacetime? If the derivation of Einstein's field equation from our model fails, all 
the other questions concerning our model, and indeed the whole model itself, will become
irrelevant.

  These questions will bring us from the quantum mechanical, statistical and 
microscopic properties of spacetime to its {\it thermodynamical} properties. Recall that
one of the starting points of our first paper in this series was Jacobson's observation 
that Einstein's field equation may be viewed, in a certain sense, as a thermodynamical
equation of state of spacetime and matter fields.\cite{nelja} In this paper our aim is to show how
the statistical properties of spacetime considered in the Section 4 of our first paper
will imply certain thermodynamical properties for spacetime, and how classical gravity
indeed follows from the thermodynamics of spacetime and matter fields. When investigating 
the thermodynamics of spacetime we consider spacetime at length scales very much larger
than the Planck scale. Because of that we are allowed to use the concepts familiar from
classical general relativity, such as metric and curvature, in our investigations. Indeed,
we saw in Section 5 of our first paper how the fundamental concepts of classical
general relativity emerge from our microscopic model of spacetime in the long distance 
limit.

  We begin our investigations in Section 2 by considering the problem of how to define the
concept of {\it heat energy} in spacetime. With the concepts of classical general 
relativity in our service, although equipped with a new interpretation, we shall focus our
attention to the objects which we shall call, for the sake of brevity and simplicity, as 
{\it acceleration surfaces}. To put it simply, an acceleration surface is a smooth, 
orientable, simply connected, spacelike two-surface of spacetime accelerating 
uniformly to the direction
of one of its spacelike unit normal fields. As a specific spacelike two-surface of 
spacetime, an acceleration surface is a long distance limit of a certain two-graph
of spacetime. Because of that an acceleration surface possesses an entropy which, according
to Eq.(1.3), is proportional to its area $A$ in the low temperature limit.

  As a simple generalization of the concept of energy as such as it is defined in 
stationary spacetimes by means of the so called Komar integrals, we define, in Section 2,
the concept of {\it heat change} of an acceleration surface. We introduce a specific 
picture of the propagation of radiation through an acceleration surface, where the 
radiation flowing through an acceleration surface picks up energy and entropy from the 
surface. Using this picture, together with our definition of heat change, we deduce an
equation which we shall call, in our model, as the "fundamental equation" of the 
thermodynamics of spacetime. That equation tells in which way radiation and acceleration
surface exchange heat energy from the point of view of an observer at rest with respect
to the acceleration surface. Most of the thermodynamical properties of spacetime obtained
in this paper, including Einstein's field equation, are simple and straightforward 
consequances of our fundamental equation. For instance, our fundamental equation, together
with Eq.(1.3) and the thermodynamical relation $\delta Q=T\,dS$, implies that
an accelerating observer will observe thermal radiation with a characteristic temperature,
which is proportional to the proper acceleration of the observer. A comparison of this 
temperature to the Unruh temperature of an accelerating observer will fix the constant 
$\alpha$ in Eq.(1.3) such that
\begin{equation}
\alpha = 2\ln 2,
\end{equation}
which implies, in the low temperature limit, that the entropy of a spacelike two-graph is,
in natural units, exactly {\it one-half} of its area. We shall also see that the Hawking 
effect is one of the consequences of Eq.(1.3) and our fundamental equation.

   In Section 3 we shall derive Einstein's field equation from our fundamental equation.
Our derivation bears some resemblance with Jacobson's derivation, and it has two steps. 
As the first step we consider masless, non-interacting radiation fields in thermal 
equilibrium with an acceleration surface of spacetime. Our fundamental equation implies
that spacetime and radiation must obey Einstein's field equation with a vanishing 
cosmological constant. A slightly different derivation is needed when the matter fields are 
massive and interacting. Again, our fundamental equation implies Einstein's field equation,
but this time with an undetermined cosmological constant. When these two derivations are 
put together, we get Einstein's field equation for general matter fields with a vanishing
cosmological constant.

   Section 4 is dedicated to the high temperature limit of our model. Among other things, 
one observes that in the high temperature limit the entropy of a spacelike two-surface
depends {\it logarithmically} on its area, instead of being proportional to the area. This 
yields radical changes to the Unruh and the Hawking effects. However, it turns out, most
curiously, that Einstein's field equation remains unchanged, no matter in which way 
entropy depends on area. This result resembles Nielsen's famous idea of 
Random Dynamics,\cite{viisi}
which states, in broad terms, that no matter what we assume about the properties of 
spacetime at the Planck scale, the low energy effects will always be the same.

   We close our discussion in Section 5 with some concluding remarks.     

\maketitle

\section{Thermodynamics of Spacetime}

\subsection{Heat and Energy}

   In the Section 5 of our first paper \cite{yksi} we found how the fundamental 
concepts of classical general
relativity arise as sort of thermodynamical quantities out of quantum spacetime. With
the concepts of classical general relativity in our service, although equipped with a 
new interpretation, we are now prepared to investigate the thermodynamics of spacetime.

   When investigating the thermodynamics of spacetime, the first task is to construct,
in the context of our model, the definitions for two fundamental concepts of 
thermodynamics. These fundamental concepts are {\it heat} and {\it temperature}. What 
do these concepts mean in quantum spacetime?

\subsubsection{Motivation: Gravitational Energy in Newtonian Gravity}

   When attempting to construct the definition of heat in quantum spacetime we can do 
nothing better than to seek for ideas and inspiration from the good old Newtonian theory of 
gravitation. This theory is based on Newton's universal law of gravitation, which states 
that point-like bodies attract each other with a gravitational force, which is directly
proportional to the masses of the bodies, and inversely proportional to the square of their
distance. This law implies that a point-like body with mass $M$ creates in its neighborhood
a gravitational field 
\begin{equation}
\vec{g}(\vec{r}) = -G\frac{M}{r^2}\hat{e}_r,
\end{equation}
where $\vec{r}$ is the position vector of the point in which the field is measured such that
the body under consideration lies at the origin of the system of coordinates. $r$ is the 
distance of that point from the body, and $\hat{e}_r$ is the unit vector parallel to 
$\vec{r}$. The gravitational field $\vec{g}(\vec{r})$ tells the {\it acceleration} an
observer at rest with respect to the body will measure, at the point $\vec{r}$, 
for all bodies in a free fall in the gravitational field created by the mass $M$. One finds
that if ${\cal S}$ is a closed, orientable two-surface, there is an interesting relationship
between the mass $M$, and the flux of the gravitational field $\vec{g}(\vec{r})$ through
that surface:
\begin{equation}
M = -\frac{1}{4\pi G}\oint_{\cal S} \vec{g}\bullet\hat{n}\,dA,
\end{equation}
where $\hat{n}$ is the outward pointing unit normal of that surface, and $dA$ is the area
element of the surface. It turns out that Eq.(2.2) holds not only for a single point-like
mass, but it holds for arbitrary mass distributions. As a generalization of Eq.(2.2) we may 
write:
\begin{equation}
M_{tot} = -\frac{1}{4\pi G}\oint_{\cal S} \vec{g}\bullet\hat{n}\,dA,
\end{equation}
where $M_{tot}$ is the total mass of the mass distribution inside the closed surface 
${\cal S}$. 
Since mass and energy are equivalent, one might expect that the right hand side of Eq.(2.3)
would provide, at least when the gravitational field is very weak, and the speeds of the
massive bodies very low, a some kind of notion of gravitational energy.

\subsubsection{Relativistic Generalization}

   The general relativistic generalization of Eq.(2.3) is obvious: In essense, we just 
replace the gravitational field $\vec{g}$, which tells the acceleration under presence
of the gravitating bodies, by the {\it proper acceleration}
\begin{equation}
a^\mu := \xi^\alpha\xi^\mu_{;\alpha}
\end{equation}
corresponding to the timelike Killing vector field $\xi^\alpha$ of spacetime. In Eq.(2.4)
the semicolon means covariant differentiation. We may define an integral 
\begin{equation}
E(V) := \frac{1}{4\pi}\oint_{\partial V} a^\mu n_\mu\,dA,
\end{equation}
where $V$ is a simply connected domain of a spacelike hypersurface of spacetime, 
$\partial V$ is its boundary, and $n_\mu$ is a spacelike unit normal vector of $\partial V$.
Another way of writing Eq.(2.5) is:
\begin{equation}
E(V) := \frac{1}{8\pi}\oint_{\partial V} \xi^{\mu;\nu}\,d\Sigma_{\mu\nu},
\end{equation}
where we have defined:
\begin{equation}
d\Sigma_{\mu\nu} := (n_\mu\xi_\nu - n_\nu\xi_\mu)\,dA.
\end{equation}
That the integrals on the right hand sides of Eqs.(2.5) and (2.6) are really the same
follows from the fact that $\xi^\mu$ obeys the Killing equation:
\begin{equation}
\xi_{\mu;\nu} + \xi_{\nu;\mu} = 0.
\end{equation}
The right hand side Eq.(2.6) is known as the {\it Komar integral} 
\cite{kuusi, seitseman, kahdeksan}, and it gives a
satisfactory definition for the concept of energy in certain stationary spacetimes.

   As an example, one may consider the Schwarzschild spacetime, where:
\begin{equation}
ds^2 = -(1-\frac{2M}{r})\,dt^2 + \frac{dr^2}{1-\frac{2M}{r}} + r^2\,d\theta^2 
+ r^2\sin^2\theta\,d\theta^2.
\end{equation}
When $r>2M$, this spacetime admits a timelike Killing vector field $\xi^\mu$ such that the
only non-zero component of this vector field is:
\begin{equation}
\xi^t \equiv 1.
\end{equation}      
On the spacelike two-sphere, where $r=constant$, the only non-zero component of $n^\mu$ is:
\begin{equation}
n^r = -(1-\frac{2M}{r})^{1/2}.
\end{equation}
One finds that the Komar integral of Eq.(2.6) becomes:
\begin{equation}
E = (1-\frac{2M}{r})^{-1/2}M.
\end{equation}
This gives the energy of the gravitational field from the pont of view of an observer at
rest with respect to the Schwarzschild coordinate $r$. $(1-(2M)/r)^{-1/2}$ is the red-shift 
factor.

    The Komar integral provides a satisfactory definition of energy in {\it stationary} 
spacetimes. How to define the concept of energy in general, non-stationary spacetimes? In 
particular, how to define the concept of {\it heat?} 

    Unfortunately, it is impossible to find a satisfactory definition of energy, or even 
energy density, in non-stationary spacetimes. (For a detailed discussion of this problem,
see Ref.\cite{yhdeksan}.). However, it might be possible to attribute meaningfully the concept of heat 
to some specific spacelike two-surfaces of spacetime. After all, we found in the Section 4 
of our first paper that spacelike two-surfaces possess entropy. If they possess entropy, then 
why should they not possess, in some sense, heat as well?

   Valuable insights into this problem are provided by the investigations we made above 
about the properties of Newtonian gravity and Komar integrals. Those investigations suggest
that when we attempt to construct a physically sensible definition of heat of spacelike
two-surfaces, the flux
\begin{equation}
\Phi := \int_{\cal S} a^\mu n_\mu\,dA
\end{equation}
of the proper acceleration vector field
\begin{equation}
a^\mu := u^\alpha u^\mu_{;\alpha}
\end{equation}
through the two-surface under consideration might play an important role \cite{kymmenen}.
 After all, both in
Eqs.(2.3) and (2.5) we calculated the flux of an acceleration vector field $a^\mu$ through a certain 
closed, spacelike two-surface. However, there is an important difference between the 
definitions (2.4) and (2.14) of the vector field $a^\mu$: In Eq.(2.4) the vector field
$a^\mu$ was defined by means of an appropriately chosen Killing vector field $\xi^\mu$, 
whereas in Eq.(2.14) $a^\mu$ is defined by means of a future pointing unit tangent vector
field $u^\mu$ of the congruence of the timelike world lines of the points of an arbitrary
spacelike two-surface of spacetime.

\subsubsection{Acceleration Surface}

    In our investigations concerning the thermodynamics of spacetime we shall focus our
attention at those smooth, orientable, simply connected spacelike two-surfaces, 
where the proper acceleration vector field $a^\mu$
of the congruence of the world lines of the points of the two-surface has the following 
properties:
\begin{subequations}
\begin{eqnarray}
\sqrt{a^\mu a_\mu} = constant := a,\\
a^\mu n_\mu = a.
\end{eqnarray}
\end{subequations}
at every point of the two-surface (In this Section we consider spacetime at macroscpic 
length scales, and therefore we may ignore its discrete substructure.). In other words, 
we shall assume that all points of the spacelike two-surface under consideration have 
all the time the same constant proper acceleration $a$, and every point of the 
two-surface is accelerated to a direction orthogonal to the two-surface. For the sake 
of brevity and simplicity we shall call such spacelike two-surfaces as {\it acceleration
surfaces}. It is easy to see that the flux of the proper acceleration vector field through 
an acceleration surface is
\begin{equation}
\Phi_{as} = aA,
\end{equation}
where $A$ is the area of the acceleration surface.

   The motivation for our definition of the concept of acceleration surface is provided 
by the fact that acceleration surfaces are very similar to the event horizons of black
holes: The surface gravity $\kappa$ is constant everywhere and all the time on a black hole
event horizon, whereas on an acceleration surface the proper acceleration $a$ is a constant.
For black hole event horizons one may meaningfully associate the concepts of heat, entropy
and temperature, and there are good hopes that the same might be done for acceleration 
surfaces as well.

\subsubsection{Properties of Acceleration Surfaces}

    A detailed investigation of the properties of acceleration surfaces has been performed
in Appendices A and B. In Appendix A a mathematically precise definition of the
concept of acceleration surface, together with some examples, is given. The main result of
Appendix A is that acceleration surface intersects orthogonally the world lines of its 
points. In other words, the vector field $u^\mu$ is orthogonal to an arbitrary tangent 
vector field of the acceleration surface everywhere and all the time on the surface. This
result is used extensively in Appendix B, where the dynamical properties of acceleration
surfaces are investigated. The main result of Appendix B is an expression for the second
proper time derivative of the area of an acceleration surface at that instant of the proper 
time $\tau$ measured along the world lines of the points of the acceleration surface, when 
the vector field $u^\mu$ has at all points of the surface the property
\begin{equation}
u^\mu_{;\nu}E^\nu_I = 0
\end{equation}
for arbitrary spacelike, orthonormal tangent vector fields $E^\mu_I$ $(I = 1, 2)$ of the surface.
If the proper time $\tau = 0$ at the instant in question, Eq.(2.17) states that the vectors 
$u^\mu$ associated with the points of the acceleration surface are parallel to each other, when 
$\tau=0$. It turns out that in this case the first proper time derivative of the area $A$ of the
acceleration surface vanishes, when $\tau=0$, i.e.
\begin{equation}
\frac{dA}{d\tau}\vert_{\tau=0} = 0,
\end{equation}
whereas the second proper time derivative takes the form:
\begin{equation}
\frac{d^2A}{d\tau^2}\vert_{\tau=0} = \int_{\mathcal{S}}(ak_n + R_{\mu\nu}u^\mu u^\nu 
- R_{\alpha\mu\nu\beta}n^\alpha n^\beta u^\mu u^\nu)\,d\mathcal{A},
\end{equation}
where $d\mathcal{A}$ is the area element on the surface, and
\begin{equation}
k_n := n^\mu_{;\nu}E^I_\mu E^\nu_I
\end{equation}
is the trace of the exterior curvature tensor induced on the surface in the direction determined
by the vector field $n^\mu$. (From this point on we shall denote the area element on the 
acceleration surface by $d\mathcal{A}$ to distinguish the area element from the infinitesimal area change
$dA$ of the acceleration surface.). So we see that if the exterior curvature tensor vanishes, i.e.
\begin{equation}
n^\mu_{;\nu}E^\nu_I = 0
\end{equation}
for all $I = 1, 2$ and at all points of the surface, when $\tau = 0$, we have:
\begin{equation}
\frac{d^2A}{d\tau^2}\vert_{\tau=0} = \int_{\mathcal{S}}(R_{\mu\nu}u^\mu u^\nu 
- R_{\alpha\mu\nu\beta}n^\alpha n^\beta u^\mu u^\nu)\,d\mathcal{A}.
\end{equation}
In other words, if both of the initial conditions (2.17) and (2.21) are satisfied by the acceleration surface,
when $\tau = 0$, then the second proper time derivative of its area depends on the Riemann 
and the Ricci tensors of spacetime only. Eq.(2.22) will play an important role in this paper.

\subsubsection{Heat Change}  

    Motivated by the similarities between black hole event horizons and acceleration 
surfaces, as well as by the properties of Komar integrals, we now define the {\it change
of heat} of an acceleration surface in terms of the differential $d\Phi_{as}$ of the flux 
$\Phi_{as}$ of the proper acceleration vector field through the acceleration surface as:
\begin{equation}
\delta Q_{as} := \frac{1}{4\pi}\,d\Phi_{as}
\end{equation}
or, in SI units:
\begin{equation}
\delta Q_{as} := \frac{c^2}{4\pi G}\,d\Phi_{as}.
\end{equation}
As such as it is, however, this is just an empty definition, and several questions arise:
What is the physical interpretation of $\delta Q_{as}$? What are its physical effects? How
to measure $\delta Q_{as}$?

\subsubsection{The Fundamental Equation}

  To find an answer to these questions, consider thermal radiation flowing through an 
acceleration surface. We parametrize the world lines of the points of the surface by means 
of the proper time $\tau$ measured along those world lines. We shall assume that the 
acceleration surface satisfies Eqs.(2.17) and (2.21), and therefore has the property:
\begin{equation}
\frac{\delta Q_{as}}{d\tau}\vert_{\tau=0} = 0,
\end{equation}
i.e. when $\tau=0$, the rate of change in the heat content of the surface is zero. 
When radiation flows through the acceleration 
surface, heat and entropy are carried through the surface, and presumably the radiation 
interacts with the surface such that its geometry is changed. For instance, the area of 
the surface may change. However, if the area of the acceleration surface changes, so 
does its heat content, and the heat delivered or absorbed by the surface contributes to
the flow $\frac{\delta Q_{rad}}{d\tau}$ of the heat $Q_{rad}$ carried by the radiation 
through the surface. As a result $\frac{\delta Q_{rad}}{d\tau}$ changes in the proper time
$\tau$, and we must have
\begin{equation}
\frac{\delta^2 Q_{rad}}{d\tau^2}\vert_{\tau=0} \ne 0,
\end{equation}
where $\frac{\delta^2 Q_{rad}}{d\tau^2}$ denotes the rate of change of 
$\frac{\delta Q_{rad}}{d\tau}$ with respect to the proper time $\tau$. The quantities 
$\frac{\delta Q_{rad}}{d\tau}$ and $\frac{\delta^2 Q_{rad}}{d\tau^2}$ have been measured
from the point of view of an observer at rest with respect to the acceleration surface.

  Conservation of energy now implies that if Eq.(2.25) holds, then the rate of increase 
in the flow of heat carried by radiation through the acceleration surface is the same as 
is the decrease in the rate of change in the heat content of the surface. In other words, 
the heat of the acceleration surface is exactly converted to the heat of the radiation,
and vice versa. A mathematical expression for this statement is:
\begin{equation}
\frac{\delta^2 Q_{rad}}{d\tau^2}\vert_{\tau=0} 
= -\frac{\delta^2 Q_{as}}{d\tau^2}\vert_{\tau=0}.
\end{equation}
In our model, this equation is the fundamental equation of the thermodynamics of spacetime.
According to the definition (2.23), Eq.(2.27) implies:
\begin{equation}
\frac{\delta^2 Q_{rad}}{d\tau^2}\vert_{\tau=0} 
= -\frac{1}{4\pi}\frac{d^2\Phi_{as}}{d\tau^2}\vert_{\tau=0}
\end{equation}
or, in SI units:
\begin{equation}
\frac{\delta^2 Q_{rad}}{d\tau^2}\vert_{\tau=0}
= -\frac{c^2}{4\pi G}\frac{d^2\Phi_{as}}{d\tau^2}\vert_{\tau=0}.
\end{equation}

\subsection{Unruh Effect}

  Consider now the possible implications of Eq.(2.27). As the first example, consider a 
very small plane, which is in a uniformly accelerating motion with a constant proper 
acceleration $a$ to the direction of its spacelike unit normal vector. Obviously, such a 
plane is an acceleration surface, and we may assume that our surface statisfies 
Eqs.(2.17) and (2.21).
Assuming that spacetime is filled with radiation in thermal equilibrium, we find that 
Eq.(2.27) implies:
\begin{equation}
\frac{\delta^2 Q_{rad}}{d\tau^2}\vert_{\tau=0} 
= -\frac{a}{4\pi}\frac{d^2 A}{d\tau^2}\vert_{\tau=0},
\end{equation}
where $A$ is the area of the plane such that $\frac{dA}{d\tau}\vert_{\tau=0}=0$. 
The first law of thermodynamics implies that
\begin{equation}
\delta Q_{rad} = T_{rad}\,dS_{rad},
\end{equation}
where $dS_{rad}$ is the amount of entropy carried by radiation out of the plane, and 
$T_{rad}$ is its temperature. Assuming that $T_{rad}$ is constant during the process, we
may write Eq.(2.30) as:
\begin{equation}
T_{rad}\frac{d^2 S_{rad}}{d\tau^2}\vert_{\tau=0} 
= -\frac{a}{4\pi}\frac{d^2 A}{d\tau^2}\vert_{\tau=0}.
\end{equation}
At this point we recall Eq.(1.3), which implies that every spacelike two-surface of 
spacetime has an entropy, which is proportional to the area of that two-surface. Using 
that equation we find that
\begin{equation}
\frac{d^2 A}{d\tau^2}\vert_{\tau=0} 
= \frac{\alpha}{\ln 2}\frac{d^2 S_{plane}}{d\tau^2}\vert_{\tau=0},
\end{equation}
where $S_{plane}$ denotes the entropy content of the plane. Hence, Eq.(2.32) takes the form:
\begin{equation}
T_{rad}\frac{d^2 S_{rad}}{d\tau^2}\vert_{\tau=0} 
= -\frac{a}{4\pi}\frac{\alpha}{\ln 2}\frac{d^2 S_{plane}}{d\tau^2}\vert_{\tau=0}.
\end{equation}
This equation allows us to associate the concept of {\it temperature} with our accelerating 
plane: When the temperatures of the radiation and the plane are equal, the entropy loss of 
the plane is equal to the entropy gain of the radiation. In other words, the entropy of the
plane is exactly converted to the entropy of the radiation. In this case the plane is in a 
thermal equilibrium with the radiation, and we have:
\begin{equation}
\frac{d^2 S_{plane}}{d\tau^2}\vert_{\tau=0} = -\frac{d^2 S_{rad}}{d\tau^2}\vert_{\tau=0},
\end{equation}
and Eq.(2.34) implies:
\begin{equation}
T_{rad} = \frac{\alpha}{\ln 2}\frac{a}{4\pi}.
\end{equation}
As one may observe, in thermal equilibrium the temperature of the radiation is proportional
to the proper acceleration $a$ of the plane.

   Now, how should we interpret this result? A natural interpretation is that an 
accelerating observer observes thermal radiation with a characteristic temperature, which 
is directly proportional to his proper acceleration. Actually, this is a well known result 
of relativistic quantum field theories, and it is known as the {\it Unruh effect} 
\cite{yksitoista}. According to 
this effect an accelerating observer observes thermal particles even when, from the point
of view of all inertial observers, there are no particles at all. The characteristic 
temperature of the thermal particles is the so called {\it Unruh temperature}
\begin{equation}
T_U := \frac{a}{2\pi}
\end{equation}
or, in SI units:
\begin{equation}
T_U := \frac{\hbar a}{2\pi k_B c}.
\end{equation}
Comparing Eqs.(2.36) and (2.37) we find that the temperature $T_{rad}$ predicted by our 
model for the thermal radiation equals to the Unruh temperature $T_U$, provided that           
\begin{equation}
\alpha = 2\ln 2.
\end{equation}
It is most gratifying that our quantum mechanical model of spacetime predicts the Unruh
effect. According to our model the Unruh effect is a direct outcome of the statistics of
spacetime.

    Eq.(2.39), together with Eq.(1.3), implies that, in the low-temperature limit, the
entropy of an arbitrary spacelike two-surface of spacetime is, in natural units,
\begin{equation}
S = \frac{1}{2}A
\end{equation}
or, in SI units,
\begin{equation}
S = \frac{1}{2}\frac{k_B c^3}{\hbar G}A.
\end{equation}
In other words, our model predicts that, in the low-temperature limit, every spacelike
two-surface of spacetime has entropy which, in natural units, is one-half of its area. 
This result is closely related to the famous {\it Bekenstein-Hawking entropy law},
which states that black hole has entropy which, in natural units, is one-quarter of its
event horizon area \cite{kaksi, kolme}. In other words, our model predicts for the entropy of a spacelike
two-surface a numerical value, which is exactly {\it twice} the numerical value of the
entropy of a black hole event horizon with the same area. At this point it should be 
noted that we would also have been able to obtain the Unruh temperature of Eq.(2.37) 
for an accelerating plane straightforwardly from the first law of thermodynamics, and an 
assumption that the plane has an entropy which is one-half of its area: It follows from 
Eqs.(2.16), (2.23) and (2.40) that if the proper acceleration $a$ on an acceleration 
surface is kept as a constant, then an infinitesimal change $\delta Q$ in its heat may
be expressed in terms of an infinitesimal change $dS$ in its entropy as:
\begin{equation}
\delta Q = \frac{a}{2\pi}\,dS,
\end{equation}
which readily implies that the temperature of the surface is the Unruh temperature $T_U$ of
Eq.(2.37). 

\subsection{Hawking Effect}

   One of the consequences of the Bekenstein-Hawking entropy law is the
{\it Hawking effect}: Black hole emits thermal particles with the characteristic
temperature
\begin{equation}
T_H := \frac{\kappa}{2\pi},
\end{equation}
which is known as the {\it Hawking temperature} \cite{kolme}. In Eq.(2.43) $\kappa$ is the surface
gravity at the horizon of the hole. For a Schwarzschild black hole with mass $M$ we have
\begin{equation}
T_H = \frac{1}{8\pi M}
\end{equation}
or, in SI units:
\begin{equation}
T_H = \frac{\hbar c^3}{8\pi G k_B}\frac{1}{M}.
\end{equation}
The Hawking temperature gives the temperature of the black hole radiation from the point
of view of a faraway observer at rest with respect to the hole, provided that the
backreaction and the backscattering effects are neglected. If the observer lies at a finite
(although very small) distance from the event horizon of the hole, Eq.(2.44) must be
corrected by the red shift factor, and we get:
\begin{equation}
T_H = \frac{1}{8\pi}(1-\frac{2M}{r})^{-1/2}\frac{1}{M}.
\end{equation}

  Consider now how Eq.(2.46), and hence the Hawking effect for the Schwarz- \newline
schild black hole,
may be obtained from our model. Our derivation of Eq.(2.46) from our model of spacetime will
also bring some light to the curious fact that the entropy of a spacelike two-surface is,
according to our model, exactly twice the entropy of a black hole event horizon with the
same area. 

  The only non-zero component of the future pointing unit tangent vector $u^\mu$ of an
observer at rest with respect to the Schwarzschild coordinates $r$ and $t$ is
\begin{equation}
u^t = (1-\frac{2M}{r})^{-1/2},
\end{equation}
and the only non-zero component of the corresponding four-acceleration $a^\mu$ is:
\begin{equation}
a^r = u^t u^r_{;t} = -\frac{M}{r^2}.
\end{equation}
It is easy to see that the $\tau=constant$ slices of the timelike hypersurfaces, where 
$r=constant(>2M)$ are acceleration surfaces.
Using Eqs.(2.11) and (2.14) we find that the flux of the vector field $a^\mu$ through that 
two-sphere is:
\begin{equation}
\Phi_{as} = 4\pi M(1-\frac{2M}{r})^{-1/2},
\end{equation}
where we have used the fact that the area of the two-sphere, where $r=constant$ is:
\begin{equation}
A = 4\pi r^2.
\end{equation}
Using Eq.(2.40) we find that in the low temperature limit the entropy of that two-sphere is:
\begin{equation}
S = 2\pi r^2.
\end{equation}

  When we obtained the Unruh effect from our model, we varied the flux $\Phi_{as}$ in Eq.(2.42)
in such a way that we kept the proper acceleration $a$ as a {\it constant}, and varied the
area $A$ only. The proper acceleration $a$ then took the role of temperature, and the area
$A$ that of entropy. We shall now use the same idea, when we obtain the Hawking effect
from our model: When varying the flux $\Phi_{as}$ of Eq.(2.43) we keep the quantity
\begin{equation}
a := a^\mu n_\mu = (1-\frac{2M}{r})^{-1/2}\frac{M}{r^2}
\end{equation}
as a constant. In other words, we consider $a$ as a function of both $M$ and $r$, and we 
require that the total differential of $a$ vanishes:
\begin{equation}
da = \frac{\partial a}{\partial M}\,dM + \frac{\partial a}{\partial r}\,dr = 0,
\end{equation}
which implies:
\begin{equation}
dM = \frac{2Mr-3M^2}{r^2-Mr}\,dr,
\end{equation}
i.e. an infinitesimal change $dM$ in the Schwarzschild mass $M$ of the hole must be 
accompained with a certain change $dr$ in the radius $r$ of the two-sphere. Using Eqs.(2.49)
and (2.54) one finds that when $a$ is kept constant, then the change $\delta Q$ in the heat 
content of the two-sphere may be written in terms of $dr$ as:
\begin{equation}
\delta Q = \frac{1}{4\pi}\,d\Phi = \frac{2M}{r}(1-\frac{2M}{r})^{-1/2}\,dr,
\end{equation}
and because Eq.(2.51) implies that the corresponding maximum  change in the entropy of the
two-sphere is
\begin{equation}
dS = 4\pi r\,dr,
\end{equation}
we find the relationship between $\delta Q$ and $dS$:
\begin{equation}
\delta Q = \frac{M}{2\pi r^2}(1-\frac{2M}{r})^{-1/2}\,dS.
\end{equation}
Therefore, according to the first law of thermodyamics, the two-sphere has a temperature
\begin{equation}
T = \frac{M}{2\pi r^2}(1-\frac{2M}{r})^{-1/2}.
\end{equation}
Hence we find that an observer with constant $r$ just outside the horizon, where $r=2M$, 
will observe that the black hole has a temperature
\begin{equation}
T_H = \frac{1}{8\pi}(1-\frac{2M}{r})^{-1/2}\frac{1}{M},
\end{equation}
which is Eq.(2.46). So we have shown that the Hawking effect is just one of the consequences
of our model in the low temperature limit.

  Let us now investigate our derivation of the Hawking effect in more details. The crucial
points in our derivation were our decisions to consider a two-sphere just outside the
horizon, instead of the horizon itself, and to keep $a^\mu n_\mu$ as a constant while 
calculating the infinitesimal change in the heat content of the two-sphere. In these
points our derivation differed from the usual derivations of the Hawking effect: In the
usual derivations one uses as the starting point the so called {\it mass formula of
black holes}, which is sometimes also known as the Smarr formula \cite{seitseman}.
For Schwarzschild black
holes this formula implies that the Schwarzschild mass $M$ of the Schwarzschild black 
hole may be written in terms of the surface gravity $\kappa$ at
the horizon, and the horizon area $A_h$ as:
\begin{equation}
M = \frac{1}{4\pi}\kappa A_h.
\end{equation}
The surface gravity $\kappa$, which may be written in terms of $M$ as:
\begin{equation}
\kappa = \frac{1}{4M},
\end{equation}
may be viewed as an analogue of the quantity $a^\mu n_\mu$ in the sense that $\kappa$ gives
the proper acceleration of an object in a free fall at the horizon from the point of view
of a faraway observer at rest with respect to the hole. From the point of view of an
observer at rest just outside the event horizon of the hole the absolute value of the
proper acceleration of objects in a free fall just outside the hole is:
\begin{equation}
a = (1-\frac{2M}{r})^{-1/2}\kappa
\end{equation}
which, according to Eqs.(2.52) and (2.61), is exactly $a^\mu n_\mu$.

   Consider now what happens when we vary the right hand side of Eq.(2.60) in such a way
that during the variation we are all the time {\it at the horizon}. In that case $\kappa$ is
{\it not constant}, but it also varies such that we have:
\begin{equation}
dM = \frac{1}{4\pi}A_h\,d\kappa + \frac{1}{4\pi}\kappa\,dA_h,
\end{equation}
where \cite{seitseman, kaksitoista}
\begin{equation}
A_h\,d\kappa = -\frac{1}{2}\,dA_h.
\end{equation}
So we get:
\begin{equation}
dM = \frac{1}{8\pi}\kappa\,dA_h,
\end{equation}
which is the first law of black hole mechanics. Identifying, as usual, $\frac{1}{4}\,dA_h$ as
the change in the entropy of the hole, and $dM$ as the change in its heat content, we get
Eq.(2.44). So we find that the reason why the black hole entropy may be thought to be 
one-quarter, instead of one-half, of the horizon area, is that when the Schwarzschild mass
$M$ of the hole decreases as a result of the black hole radiance, the event horizon of the hole
shrinks such that the surface gravity $\kappa$ changes in the manner described in Eq.(2.64).
If the right hand side of Eq.(2.60) were varied in such a way that $\kappa$ is kept 
as a constant, then the entropy change corresponding to the area change $dA_h$ at the
temperature $T_H$ of Eq.(2.46) would be one-half, instead of one-quarter of $dA_h$. 

\maketitle 

\section{Classical Limit: Einstein's Field Equation}

 We saw in the previous Section, much to our satisfaction, that our quantum mechanical model
of spacetime reproduces, in the low temperature limit, both the Unruh and the Hawking
effects. In other words, our model reproduces the well known semiclassical effects of 
gravity. 

   A really interesting question, and indeed the crucial test for our model, is whether the 
model implies, in the classical limit, Einstein's field equation. If it does, then we may 
say that Einstein's classical general relativity with all of its predictions is just one of
the consequences of our model.

   In this Section we show that Einstein's field equation indeed follows from our model in 
the classical limit. Our derivation will be based on the thermodynamical properties of 
spacetime, which were considered in the previous Section. It turns out that Einstein's field
equation is a simple and straightforward consequence of Eq.(2.27), the fundamental equation
of the thermodynamics of spacetime in our model. As in the previous Section, we consider
spacetime at length scales very much larger than the Planck length scale. At these length
scales we may consider spacetime, in effect, as a smooth (pseudo-) Riemannian manifold.

\subsection{Boost Energy Flow}

   As the first step in the derivation of Einstein's field equation from Eq.(2.27) let us 
consider the flow of boost energy through an acceleration surface. In general, the amount of boost energy 
flown during a unit proper time, or {\it boost energy flow} through an arbitrary spacelike 
two-surface $\mathcal{S}$ to the direction of a spacelike unit normal $n^\mu$ of the surface is,
for general matter fields:
\begin{equation}
\frac{dE_b}{d\tau} = \int_{\mathcal{S}}T_{\mu\nu}u^\mu n^\nu\,d\mathcal{A}
\end{equation}
where, as in the previous Section, $u^\mu$ is the future directed unit tangent vector field 
of the congruence of the world lines of the points of the surface. $T_{\mu\nu}$ is the
energy momentum stress tensor of the matter fields, and $d\mathcal{A}$ is the area element
on the surface. In what follows, we shall assume that $\mathcal{S}$ is an acceleration surface
which obeys the initial conditions (2.17) and (2.21), and we consider the situation at the moment
$\tau = 0$ of the proper time $\tau$ measured along the world lines of the surface. Since Eq.(2.17)
implies that $\frac{dA}{d\tau}\vert_{\tau=0} = 0$, we find that the rate of change in the boost energy 
flow is, when $\tau=0$:
\begin{equation}
\frac{d^2E_b}{d\tau^2}\vert_{\tau=0} = \int_{\mathcal{S}}\frac{d}{d\tau}(T_{\mu\nu}u^\mu n^\nu)\,d\mathcal{A}
 = \int_{\mathcal{S}}u^\alpha(T_{\mu\nu}u^\mu n^\nu)_{;\alpha}\,d\mathcal{A},
\end{equation}
where we have used the chain rule. We have been allowed to replace the ordinary partial derivatives by the 
covariant ones, because the expression inside the brackets is a scalar. Since it follows from the 
considerations made in the Appendix A that the vector fields $u^\mu$ and $n^\mu$ have the properties:
\begin{subequations}
\begin{eqnarray}
u^\alpha u^\mu_{;\alpha} &=& an^\mu,\\
u^\alpha n^\mu_{;\alpha} &=& au^\mu,
\end{eqnarray}
\end{subequations}
we find, by means of the product rule of covariant differentiation:
\begin{equation}
\frac{d^2E_b}{d\tau^2}\vert_{\tau=0} = \frac{d^2E_{b,t}}{d\tau^2}\vert_{\tau=0} 
+ \frac{d^2E_{b,a}}{d\tau^2}\vert_{\tau=0},
\end{equation}
where we have defined:
\begin{subequations}
\begin{eqnarray}
\frac{d^2E_{b,t}}{d\tau^2}\vert_{\tau=0} &:=& \int_{\mathcal{S}} 
T_{\mu\nu;\alpha}u^\alpha u^\mu u^\nu\,d\mathcal{A},\\
\frac{d^2E_{b,a}}{d\tau^2}\vert_{\tau=0} &:=& a\int_{\mathcal{S}}
T_{\mu\nu}(u^\mu u^\nu + n^\mu n^\nu)\,d\mathcal{A},
\end{eqnarray}
\end{subequations}
The presence of the first term on the right hand side of Eq.(3.4) is simply caused by the fact that
the acceleration surface propagates in spacetime, and the tensor $T^{\mu\nu}$ may be different in
different points of spacetime, whereas the second term is caused by the mere acceleration of the
surface: If the speed of a surface with respect to the matter fields changes, so does the boost 
energy flow through the surface. In what follows, we shall always assume that the proper 
acceleration $a$ of the acceleration surface is so large that the second term vastly exceeds the 
first term, and we may neglect the first term.

\subsection{Einstein's Field Equation for Massless, Non-Interacting Radiation}

    Eq.(2.27), the fundamental equation of the thermodynamics of spacetime, was originally written for
{\it radiation} interacting with an acceleration surface. Because of that, let us first consider a 
special case, where matter consists of massless, non-interacting radiation in thermal equilibrium. 
A typical example of this kind of radiation is, of course, the electromagnetic radiation. The energy
density of massless, non-interacting radiation in thermal equilibrium is, in the rest frame of 
our acceleration surface,
\begin{equation}
\rho = T_{\mu\nu}u^\mu u^\nu,
\end{equation}
its pressure is
\begin{equation}
p = \frac{1}{3}\rho = T_{\mu\nu}n^\mu n^\nu,
\end{equation}
and the energy momentum stress tensor is {\it traceless}. In other words,
\begin{equation}
T^\alpha_{\,\,\alpha} = 0.
\end{equation}
Using Eq.(3.5b) we therefore find that
\begin{equation}
\frac{d^2E_{b,a}}{d\tau^2}\vert_{\tau=0} = \frac{4}{3}a\int_{\mathcal{S}}\rho\,d\mathcal{A}
 = \frac{4}{3}a\int_{\mathcal{S}} T_{\mu\nu}u^\mu u^\nu\,d\mathcal{A}.
\end{equation} 

   It is easy to see that the right hand side of Eq.(3.9) gives the rate of change in 
the flow of {\it heat} through our accelerating plane, provided that all change in the flow
of boost energy is, in effect, caused by the mere acceleration of the plane. This
important conclusion follows from the well known fact that the entropy density (entropy
per unit volume) of the electromagnetic (or any massless, non-interacting) radiation in
thermal equilibrium is \cite{neljatoista}
\begin{equation}
s_{rad} = \frac{1}{T_{rad}}\frac{4}{3}\rho,
\end{equation}
where $T_{rad}$ is the absolute temperature of the radiation. One easily finds, by using the
first law of thermodynamics, that the rate of change in the flow of heat through our 
accelerating plane is
\begin{equation}
\frac{\delta^2 Q_{rad}}{d\tau^2}\vert_{\tau=0} = \frac{4}{3}a\int_{\mathcal{S}}\rho\,d\mathcal{A},
\end{equation}
which is exactly Eq.(3.9).

    It is now very easy to obtain Einstein's field equation. We just use Eq.(2.30) which, in
turn, follows from Eq.(2.27). We shall assume that our plane is initially at rest
with respect to the radiation, i.e.
\begin{equation}
T_{\mu\nu}u^\mu n^\nu = 0
\end{equation}
at every point of the surface,when $\tau = 0$. In that case there is no net flow of radiation 
through the plane, and because radiation is 
assumed to be in thermal equilibrium, and therefore homogeneous and isotropic, spacetime 
expands and contracts in the same ways in all spatial directions. This implies that when 
$\tau=0$, then at every point of the surface:
\begin{equation}
R_{\alpha\mu\nu\beta}E^\alpha_{(1)}E^\beta_{(1)}u^\mu u^\nu 
= R_{\alpha\mu\nu\beta}E^\alpha_{(2)}E^\beta_{(2)}u^\mu u^\nu 
= R_{\alpha\mu\nu\beta}n^\alpha n^\beta u^\mu u^\nu
\end{equation}
for arbitrary spacelike, orthonormal tangent vector fields $E^\mu_{(1)}$ and $E^\mu_{(2)}$
of the surface. Assuming that Eq.(2.21) holds, we find, using Eq.(2.22), that the second proper time
derivative of the area of the acceleration surface takes, in this special case, the form:
\begin{equation}
\frac{d^2A}{d\tau^2}\vert_{\tau=0} = \frac{2}{3}\int_{\mathcal{S}} 
R_{\mu\nu}u^\mu u^\nu\,d\mathcal{A}.
\end{equation}
Hence we get, using Eq.(3.9) for the left hand side, and Eq.(3.14) for the right hand side
of Eq.(2.30):
\begin{equation}
\frac{4}{3}a\int_{\mathcal{S}}T_{\mu\nu}u^\mu u^\nu\,d\mathcal{A} = -\frac{a}{6\pi}\int_{\mathcal{S}}
R_{\mu\nu}u^\mu u^\nu\,d\mathcal{A}.
\end{equation}
Since the acceleration surface $\mathcal{S}$, as well as the timelike vector field $u^\mu$,
are arbitrary, we must have:
\begin{equation}
R_{\mu\nu} = -8\pi T_{\mu\nu},
\end{equation}
which is exactly Einstein's field equation 
\begin{equation}
R_{\mu\nu} = -8\pi(T_{\mu\nu} - \frac{1}{2}g_{\mu\nu}T^\alpha_\alpha),
\end{equation}
or
\begin{equation}
R_{\mu\nu} - \frac{1}{2}g_{\mu\nu}R = -8\pi T_{\mu\nu}
\end{equation}
in the special case, where Eq.(3.12) holds, i.e. the energy momentum stress tensor is
traceless. In other words, we have obtained Einstein's field equation in the special 
case, where matter consists of massless, non-interacting radiation in thermal equilibrium.

   Our derivation of Einstein's field equation was entirely based on Eq.(2.27), the 
fundamental equation of the thermodynamics of spacetime in our model. That equation
states just the conservation of energy, when the matter flowing through an acceleration
surface consists of radiation, and the energy it carries of heat only. Actually, we did
not even need the result that the entropy of a spacelike two-surface of spacetime is, in
natural units, exactly one-half of its area in the low temperature limit. Our succesful 
derivation of Einstein's field equation provides a strong argument for the validity of
Eq.(2.27), as well as for the idea that one may meaningfully associate the concept
of heat with the acceleration surfaces. 

\subsection{Einstein's Field Equation for General Matter Fields}

   After obtaining Einstein's field equation, when matter consists of massless, 
non-interacting radiation in thermal equilibrium, the next challenge is to derive that equation
for general matter fields. In doing so, however, we meet with some difficulties, because
Eq.(2.27) is assumed to hold for {\it radiation} only. Moreover, the rate of change in the 
boost energy flow through the plane should be, in effect, the rate of change in the flow
of {\it heat}. The problem is that for general matter fields other forms of energy, except heat
(mass-energy, for instance) flow through the plane, and therefore it seems that we cannot
use the same kind of reasoning as we did above.

   These issues were investigated in details in Ref.\cite{kolmetoista}. The object of study
in Ref.\cite{kolmetoista} was, instead of an acceleration surface, an infinitesimal, 
accelerating, spacelike two-plane. The idea was to make the two-plane to move, with
respect to the matter fields, with a velocity very close to that of light, which means
that the particles of matter fields move, in the rest frame of the plane, with enormous 
velocities through the plane. It was shown that in this limit we may consider arbitrary
matter, in the rest frame of the plane, in effect, as a gas of non-interacting massless 
particles, regardless of the kind of matter we happen to have. More precisely, it was shown 
that in the high speed limit the components of the energy momentum stress tensor 
$T^{\mu\nu}$ of arbitrary matter become, in the rest frame of the plane, identical
to those of a gas of massless, non-interacting particles. In other words, all matter behaves,
as far as we are interested in its energy momentum stress tensor only, like massless, 
non-interacting radiation, provided that we move fast enough with respect to the matter. 
One may also show that in the high speed limit Eq.(3.5b) gives exactly the rate of change 
in the flow of {\it heat} through an accelerating plane for arbitrary matter. As whole,
therefore, we observe that Eq.(2.27) may be applied as such for general matter fields at 
least in the special case, where our acceleration surface is an infinitesimal, spacelike
two-plane moving with a very high speed with respect to the matter fields.

   We shall now utilize these ideas when attempting to obtain Einstein's field equation 
from Eq.(2.27) for general matter fields. As the first step we shall assume that all 
components of the Riemann and the Ricci tensors of spacetime, as well as the components
of the energy momentum stress tensor $T^{\mu\nu}$ of the matter fields, are fixed and finite 
at the points of our acceleration surface in the rest frame of the surface when $\tau = 0$. 
More precisely, we shall assume that if we project these tensors along the vectors $u^\mu$,
$n^\mu$ and $E^\mu_I$ $(I = 1, 2)$, we get fixed and finite numbers at every point of the 
surface.

   As the second step we Lorentz boost every point on our acceleration surface to the direction
of the vector $-n^\mu$. In that case the spacelike unit tangent vector fields $E^\mu_I$ of the
surface will preserve invariant, but the vector fields $u^\mu$ and $n^\mu$ will transform
to the vector fields $u'^\mu$ and $n'^\mu$ such that:
\begin{subequations}
\begin{eqnarray}
u'^\mu &=& u^\mu\,\cosh\phi - n^\mu\,\sinh\phi,\\
n'^\mu &=& u^\mu\,\sinh\phi - n^\mu\,\cosh\phi.
\end{eqnarray}
\end{subequations}
In these equations 
\begin{equation}
\phi := \sinh^{-1}(\frac{v}{\sqrt{1-v^2}})
\end{equation}
is the boost angle, or rapidity, and $v$ is the speed of the boosted point of the surface
with respect to the original point. Introducing a parameter
\begin{equation}
\epsilon := \frac{1-v}{1+v}
\end{equation}
we find that Eq.(3.19) may be written as:
\begin{subequations}
\begin{eqnarray}
u'^\mu &=& \frac{1}{2}(\frac{k^\mu}{\sqrt{\epsilon}} + \sqrt{\epsilon}\,l^\mu),\\
n'^\mu &=& \frac{1}{2}(\frac{k^\mu}{\sqrt{\epsilon}} - \sqrt{\epsilon}\,l^\mu),
\end{eqnarray}
\end{subequations}
where we have defined the future directed null vector fields $k^\mu$ and $l^\mu$ such that:
\begin{subequations}
\begin{eqnarray}
k^\mu &:=& u^\mu - n^\mu,\\
l^\mu &:=& u^\mu + n^\mu.
\end{eqnarray}
\end{subequations}
In a given point of our acceleration surface $k^\mu$ generates the past, and $l^\mu$ the future
local Rindler horizon of that point. As one may observe form Eq.(3.21), the parameter $\epsilon$ 
goes to zero, when $v$ goes to one, the speed of light in the natural units. Hence the limit, 
where our acceleration surface moves with an enormous velocity with respect to the matter 
fields corresponds to the limit, where $\epsilon\longrightarrow 0$.

   To investigate the high speed limit we replace the vector fields $u^\mu$ and $n^\mu$ in 
Eqs.(2.22) and (3.5b) by the vector fields $u'^\mu$ and $n'^\mu$ written by means of the
null vector fields $k^\mu$ and $l^\mu$, and the parameter $\epsilon$. Using the symmetry 
properties of the Riemann and the Ricci tensors we find:
\begin{equation}
\begin{split}
R_{\mu\nu}u'^\mu u'^\nu - R_{\alpha\mu\nu\beta}n'^\alpha n'^\beta u'^\mu u'^\nu
&= \frac{1}{4\epsilon}R_{\mu\nu}k^\mu k^\nu + \frac{1}{2}R_{\mu\nu}k^\mu l^\nu 
- \frac{1}{4}R_{\alpha\mu\nu\beta}k^\alpha k^\beta l^\mu l^\nu\\
&\quad + \frac{\epsilon}{4}R_{\mu\nu}l^\mu l^\nu.
\end{split}
\end{equation}
As one may observe, the first term on the right hand side of this equation will dominate 
in the high speed limit, where $\epsilon\longrightarrow 0$. Hence we may write Eq.(2.22),
for very small $\epsilon$, in the form:
\begin{equation}
\frac{d^2A}{d\tau^2}\vert_{\tau=0} = \frac{1}{4\epsilon}\int_{\mathcal{S}} 
R_{\mu\nu}k^\mu k^\nu\,d\mathcal{A} + \mathcal{O}(1),
\end{equation}
where $\mathcal{O}(1)$ denotes the terms, which are of the order $\epsilon^0$, or higher. Correspondingly,
we may write Eq.(3.5b) in the form:
\begin{equation}
\frac{d^2E_{b,a}}{d\tau^2}\vert_{\tau=0} = \frac{a}{2\epsilon}\int_{\mathcal{S}}
T_{\mu\nu}k^\mu k^\nu\,d\mathcal{A} + \mathcal{O}(\epsilon),
\end{equation}
where $\mathcal{O}(\epsilon)$ denotes the terms, which are of the order $\epsilon^1$,
or higher. 

   We may identify $\frac{d^2E_{b,a}}{d\tau^2}\vert_{\tau=0}$ as 
$\frac{\delta^2Q_{rad}}{d\tau^2}\vert_{\tau=0}$, when $\epsilon\longrightarrow 0$. So we
find that Eq.(2.30) implies, in the limit, where $\epsilon\longrightarrow 0$:
\begin{equation}
\frac{a}{2}\int_{\mathcal{S}}T_{\mu\nu}k^\mu k^\nu\,d\mathcal{A} 
= -\frac{a}{16\pi}\int_{\mathcal{S}}R_{\mu\nu}k^\mu k^\nu\,d\mathcal{A},
\end{equation}
and since our acceleration surface $\mathcal{S}$ is arbitrary, we have:
\begin{equation}
R_{\mu\nu}k^\mu k^\nu = -8\pi T_{\mu\nu}k^\mu k^\nu
\end{equation}
for general matter fields, at every point of our acceleration surface. Because $k^\mu$ may be
chosen to be an arbitrary, future directed null vector field, we must have:
\begin{equation}
R_{\mu\nu} + fg_{\mu\nu} = -8\pi T_{\mu\nu},
\end{equation}
where $f$ is some function of the spacetime coordinates. It follows from the Bianchi identity
\begin{equation}
(R^\mu_{\,\,\,\nu} - \frac{1}{2}R\delta^\mu_\nu)_{;\mu} = 0,
\end{equation}
that
\begin{equation}
f = -\frac{1}{2}R + \Lambda
\end{equation}
for some constant $\Lambda$, and hence we arrive at the equation
\begin{equation}
R_{\mu\nu} - \frac{1}{2}g_{\mu\nu}R + \Lambda g_{\mu\nu} = -8\pi T_{\mu\nu},
\end{equation}
which is Einstein's field equation with the cosmological constant $\Lambda$. We have thus 
achieved our goal: We have obtained Einstein's field equation for general matter fields in the
long distance limit.

\subsection{The Cosmological Constant}

   Our derivation of Einstein's field equation was entirely based on Eq.(2.27), the fundamental 
equation of the thermodynamics of spacetime in our model. Hence our derivation provides support
for the view, advocated by Jacobson and others, \cite{nelja} that Einstein's field equation is actually a
thermodynamical equation of state of spacetime and matter fields. However, there are some 
important differences between our derivation, and the other thermodynamical derivations of
Einstein's field equation presented so far. For instance, all of the previously expressed
thermodynamical derivations have left the cosmological constant $\Lambda$ completely 
unspecified, whereas our derivation provides for the cosmological constant a precise numerical
value: Since Eq.(3.32) holds for general matter fields, and Eq.(3.18) for massless, 
non-interacting radiation, Eq.(3.32) should reduce to Eq.(3.18) when matter consists of 
massless, non-interacting thermal radiation only. Obviously, this is not the case, unless the 
cosmological will vanish. In other words, the cosmological constant must be exactly zero:
\begin{equation}
\Lambda = 0.
\end{equation}
So we see that Eq.(2.27) makes a precise prediction, which is consistent with the present 
observations: The cosmological constant, although not necessarily exactly zero, must 
nevertheless be very small. \cite{viisitoista}    

  The most important feature of our derivation of Einstein's field equation from 
thermodynamical considerations is that it proves the validity of Eq.(2.27), the fundamental
equation of the thermodynamics of spacetime in our model. Indeed, Eq.(2.27) implies 
Einstein's field equation with a vanishing cosmological constant, and vice versa. Without 
Eq.(2.27) we would not have been able to derive, for instance, the Unruh and the Hawking 
effects from our model. Since these effects are certainly of quantum mechanical origin, 
Eq.(2.27), whose validity was proved in this Section, provides a necessary bridge between
the quantum mechanics and the thermodynamics of spacetime. Hence our thermodynamical 
derivation of Einstein's field equation is an absolutely essential element in our 
discussion of the quantum mechanical properties of spacetime.

\maketitle

\section{The High Temperature Limit}

   It was shown in Section 4 of our first paper that in the {\it high temperature limit}, where most microscopic
quantum black holes lying on a two-dimensional subgraph of spacetime are in highly excited states, the entropy of
that subgraph may be written effectively in terms of its area $A$ and the number $N$ of the holes as:
\begin{equation}
S = N\ln A.
\end{equation}
In other words, when the microscopic quantum black holes in the spacetime region under 
consideration are, in average, in highly excited states, the entropy $S$ is not proportional
to the area $A$, but it depends {\it logarithmically} on $A$. It has been speculated for a 
long time by several authors that there might be, in addition to a simple proportionality,
a logarithmic dependence between area and entropy \cite{kuusitoista}.
Our model implies that there indeed 
exists such a dependence, and this logarithmic dependence dominates in a certain limit. 

  Consider now the physical consequences of Eq.(4.1). We find that between the 
infinitesimal changes in the entropy $S$ and the area $A$ there is a relationship:
\begin{equation}
dS = \frac{N}{A}\,dA.
\end{equation}
As it was discussed in Section 2, the area of a given spacelike two-surface of spacetime
may change, for instance, when radiation goes through that two-surface. It follows from 
the results and the definitions of Section 2 that if we have an acceleration surface, then 
the change in the {\it heat content} of the surface is
\begin{equation}
\delta Q = \frac{a}{4\pi}\,dA,
\end{equation}
provided that $a$ is kept as a constant during the variation of $Q$. Because, according to 
the first law of thermodynamics,
\begin{equation}
\delta Q = T\,dS,
\end{equation}
Eqs.(4.2) and (4.3) imply:
\begin{equation}
NT = \frac{a}{4\pi}A.
\end{equation}
Using Eq.(4.3) we get:
\begin{equation}
\delta Q = N\,dT
\end{equation}
or, in SI units:
\begin{equation}
\delta Q = Nk_B\,dT.
\end{equation}

   To some extent, Eq.(4.7) may be used as a consistency check of our model. In the high temperature
 limit the thermal 
fluctuations of the areas of the black holes on two-dimensional subgraph become so large that the
quantum mechanical discreteness of the area spectrum is, in effect, washed out into 
continuum, and classical statistical mechanics may be applied. It is a general feature of
almost any system that in a sufficiently high temperature the relationship between the 
infinitesimal changes in the heat $Q$, and in the absolute temperature $T$ of
the system is of the form:
\begin{equation}
\delta Q = \gamma Nk_B\,dT.
\end{equation}
In this equation, $N$ is the number of the constituents (atoms or molecules) of the system,
and $\gamma$ is a pure number of order one, which depends on the physical properties (the 
number of independent degrees of freedom, etc.) of the system. For instance, in sufficiently
high temperatures most solids obey the Dulong-Petit law \cite{neljatoista}:
\begin{equation}
\delta Q = 3Nk_B\,dT,
\end{equation}
where $N$ is the number of molecules in the solid. From Eq.(4.7) we observe that this 
general feature is also possessed by our spacetime model. Eq.(4.7) is exactly what one
expects in the high temperature limit on grounds of general statistical arguments.

   According to Eq.(4.5) the absolute temperature $T$ measured by an observer at rest
with respect to an accelerating, spacelike two-surface depends, in the high temperature 
limit, on the proper acceleration $a$, area $A$, and the number $N$ of the microscopic 
quantum black holes on the surface such that:
\begin{equation}
T = \frac{a}{4\pi N}A
\end{equation}
or, in SI units:
\begin{equation}
T = \frac{c^2 a}{4\pi Nk_BG}A.
\end{equation}
Our derivation of Eq.(4.11) was analogous to the derivation of the expression for the 
Unruh temperature $T_U$ in Eq.(2.38). Indeed, Eq.(4.11) represents, in the high temperature
limit, the minimum temperature an accelerating observer may 
measure. In other words, Eq.(4.29) gives the Unruh temperature measured by an accelerating
observer when most of the microscopic black holes constituting the spacetime region under 
consideration are in highly excited states. As one may observe, Eq.(4.29) is radically different from
Eq.(2.38): In the high temperature limit the Unruh temperature is not constant for constant $a$, 
but it is directly proportional to the ratio $A/N$.

  Eq. (4.1) has interesting consequences for Hawking radiation: If we substitute for $A$ 
$4\pi r^2$, which gives the area of a two-sphere 
surrounding the event horizon of a Schwarzschild black hole, and for $a$ the expression
the expression for $a^\mu n_\mu$ in Eq.(2.52), we find that in the high temperature limit
the Hawking temperature of a Schwarzschild black hole measured by an observer just outside
of the event horizon of the hole is:
\begin{equation}
T_H = (1 - \frac{2M}{r})^{-1/2}\frac{M}{N}
\end{equation}
or, in SI units:
\begin{equation}
T_H = (1 - \frac{2GM}{c^2 r})^{-1/2}\frac{Mc^2}{Nk_B}.
\end{equation}
As one may observe, for warm enough black holes the Hawking temperature will no more be
inversely, but {\it directly} proportional to the Schwarzschild mass $M$ of the hole. 

   It is interesting that neither of the Eqs.(4.11) and (4.13) involve the Planck constant
$\hbar$. This is something one might expect: In very high temperatures the thermal 
fluctuations in the horizon area eigenvalues of the microscopic quantum black holes become
so large that the discrete area spectrum predicted by quantum mechanics is, in effect, 
washed out into continuum. In this limit quantum effects on the statistics of spacetime
may be ignored, and classical statistics may be applied. In contrast to the high temperature
limit, where quantum effects become negligible, in the {\it low temperature limit} the
quantum effects of spacetime play an essential role. In this sense we may say that, contrary
to the common beliefs, spacetime behaves quantum mechanically in low temperatures, and 
classically in high temperatures. This explains the presence of the Planck constant 
$\hbar$ in the low temperature formulas (2.38) and (2.45), and its absence in the high 
temperature formulas (4.11) and (4.13): Eqs.(2.38) and (2.45) are quantum mechanical,
whereas Eqs.(4.11) and (4.13) are classical. 

  Which form will Einstein's field equation take in the high temperature limit? To answer 
this question, recall that in Section 3 we obtained Einstein's field equation by means of 
our fundamental thermodynamical equation (2.27) only. Nowhere in our derivation did we 
use any explicit relationship between the area and the entropy of a spacelike two-surface
of spacetime. Since the temperature of spacetime has effects on this relationship only,
we find that if we assume that Eq.(2.27) holds as such for all temperatures, no matter
whether those temperatures are low or high, Einstein's field equation is independent of 
that temperature. In other words, Einstein's field equation takes in high temperatures 
exactly the same form as it does in low temperatures. Since Einstein's field equation
is independent of the relationship between area and entropy, it is independent of the 
precise microscopic physics of spacetime. This startling conclusion reminds us of Nielsen's
famous idea of Random Dynamics \cite{viisi}. According to Nielsen's idea nature behaves 
in such a 
way that no matter what we assume of its behavior at the Planck energy scales, the 
consequences of those assumptions will always produce, as sort of statistical averages, the
well known laws of physics in the low energy limit. Indeed, our model is in harmony with 
this idea: The high energy effects such as the Unruh and the Hawking effects, depend 
crucially on the microphysics of spacetime, whereas the low energy effects, such as 
classical gravity, are independent of that microphysics.

\maketitle

\section{Concluding Remarks}

  In this second part of our series of two papers we have considered the thermodynamical 
properties of the spacetime model introduced in the first part of our series. As in our 
first paper, \cite{yksi} Planck size quantum black holes were taken to be the fundamental constituents
of spacetime. Spacetime was assumed to be a graph, where black holes lie on the vertices.

  Our thermodynamical investigations were based on the concept of acceleration surface.
Acceleration surface may be defined as a smooth, orientable, simply connected, spacelike 
two-surface of 
spacetime, whose every point accelerates uniformly to the direction of a spacelike normal
of the surface. For acceleration surfaces we introduced the concept of heat change, which
is proportional to the flux of the proper acceleration vector field of the congruence of
the world lines of the points of the surface through the surface, and an equation, which
we called, in our model, as the "fundamental equation" of the thermodynamics of spacetime.
In broad terms, our fundamental equation tells in which way acceleration surface and 
radiation flowing through the surface exchange heat with each other. By means of our 
fundamental equation and the result, found in our first paper, that every spacelike 
two-surface of spacetime posesses an entropy which, in the low temperature limit, is 
proportional to its area, we derived the Unruh and the Hawking effects from our model. 
We also found that Einstein's field equation with a vanishing cosmological constant is a
straightforward consequence of our fundamental equation. Our derivation of Einstein's
field equation from the fundamental equation involved two steps. As the first step we 
derived Einstein's field equation with a vanishing cosmological constant for massless, 
non-interacting radiation (electromagnetic radiation, for instance) in thermal equilibrium.
A slightly different derivation was needed for general matter fields. That derivation 
implied Einstein's field equation with an unspecified cosmological constant. When those two
derivations were put together, we got Einstein's field equation with a vanishing 
cosmological constant for general matter fields. 

  In addition to the low temperature limit, where most of the Planck size quantum black 
holes constituting spacetime were assumed to be close to the ground state, we also 
considered the high temperature limit of our model, where the Planck size quantum black
holes were assumed to be in highly excited states. In this limit one finds that the entropy
of spacelike two-surfaces of spacetime, instead of being proportional to the area, depends
logarithmically on the area. Although this yields radical changes to the Unruh and the 
Hawking effects in the high temperature limit, we found that Einstein's field equation
nevertheless remains the same.

   Taken as a whole, our two papers may be viewed as an attempt to probe a possibility to 
construct an entirely novel approach to quantum gravity. Instead of trying to quantize 
general relativity as if it were an ordinary field theory in the same sense as, 
for instance, classical electromagnetism, one postulates certain microscopic, 
quantum mechanical properties for the fundamental constituents of space and time. Using
these postulates one attempts to obtain the "hard facts" of gravitational physics as 
such as we know them today, in the long distance and the thermodynamical limit. Of course,
one also hopes to be able to produce some new, observationally testable predictions. The
results of our two papers encourage one to think that such an approach may indeed be  
possible: In addition to the postulates posed for the microscopic quantum black holes 
acting as the fundamental building blocks of space and time, we used our 
"fundamental equation" only in the derivation of the Unruh and the Hawking effects, 
together with Einstein's field equation, from our model.

  Actually, our "fundamental equation" is very natural: In effect, it is just an attempt 
to generalize the principle of energy conservation from flat to curved spacetime in the 
special case where matter consists of massless, non-interacting radiation only. One may
expect that an equation of that kind is always needed when one attempts to obtain the 
classical and the semiclassical effects of gravity from the postulates posed for the
fundamental constituents of spacetime. After all, the Unruh and the Hawking effects, as 
well as Einstein's field equation, result from the thermodynamical properties of spacetime.
The thermodynamical properties of any system, in turn, are never consequences of the 
microphysics of the system alone, but they also follow from the laws of thermodynamics. 
Our fundamental equation tells in which way the laws of thermodynamics should be applied,
in certain special cases, in curved spacetime interacting with matter fields.

    In some respects, the results given by our model may be viewed as a confirmation
of the validity of an idea, first expressed by 't Hooft, that in gravitational physics the observational degrees
of freedom may be described as if they were variables defined on a two-dimensional lattice evolving in time. 
\cite{nobelisti}
Indeed, we found that Einstein's field equation, as well as the Unruh and the Hawking effects, follow from the
properties of an acceleration surface which, in the microscopic level, is a specific two-dimensional lattice with
microscopic black holes on its vertices. Most likely, our model is still far away from a decent proposal for a 
proper quantum theory of gravity. Nevertheless, the message it conveys is clear: Quantization of gravity 
may well be more simple than it has generally been thought so far.

\appendix

\section{Acceleration Surface: Definition and Examples}

Even though the physical meaning of the concept of acceleration surface as a 
straightforward generalization of the black hole event horizon is fairly simple,
its mathematically precise definition is rather tricky. The definition involves 
two concepts which we shall call, for the sake of simplicity, as {\it acceleration
curve} and {\it acceleration congruence.}

\subsection{Acceleration Curve}

\subsubsection{The Definition of Acceleration Curve}

  By {\it acceleration curve} we mean a smooth, timelike, future directed curve such that the 
norm
\begin{equation}
\vert\vert a^\mu \vert \vert := \sqrt{a^\mu a_\mu} :=a
\end{equation}
of the proper acceleration vector $a^\mu$ is constant at every point of the curve. In 
general, if we parametrize an acceleration curve by the proper time $\tau$ measured along
the curve, the proper acceleration vector is
\begin{equation}
a^\mu = \frac{du^\mu}{d\tau} + \Gamma^\mu_{\alpha\beta}u^\alpha u^\beta,
\end{equation}
where
\begin{equation}
u^\mu = u^\mu(\tau) := \frac{dx^\mu}{d\tau}
\end{equation}
is the future directed unit tangent vector of the curve. A special case of an acceleration 
curve is the one, where $a$ vanishes identically. In that case all components of $a^\mu$ are 
identically zero, and $u^\mu$ satisfies the geodesic equation
\begin{equation}
\frac{du^\mu}{d\tau} + \Gamma^\mu_{\alpha\beta}u^\alpha u^\beta = 0.
\end{equation}
In other words, the acceleration curve, in this special case, is a timelike geodesic of 
spacetime.

\subsubsection{Examples of Acceleration Curves}

  The simplest possible non-trivial example of an acceleration curve is the world line of 
a uniformly accelerating observer in flat Minkowski spacetime. If the observer is accelerated, 
in space, uniformly to the direction of the positive $x$-axis with a constant proper acceleration
$a$, we may write the equation of the world line of that observer in a parametrized form as:
\begin{subequations}
\begin{eqnarray}
t(\tau) &=& \frac{1}{a}\sinh(a\tau),\\
x(\tau) &=& \frac{1}{a}\cosh(a\tau),
\end{eqnarray}
\end{subequations}
where $t$ and $x$ are flat Minkowski coordinates. The non-zero components of the vector $u^\mu(\tau)$
are:
\begin{subequations}
\begin{eqnarray}
u^0(\tau) &=& \frac{dt(\tau)}{d\tau} = \cosh(a\tau),\\
u^1(\tau) &=& \frac{dx(\tau)}{d\tau} = \sinh(a\tau),
\end{eqnarray}
\end{subequations}
and because $\Gamma^\mu_{\alpha\beta}\equiv 0$ in flat Minkowski spacetime equipped with Minkowski 
coordinates, the non-zero components of $a^\mu(\tau)$ are:
\begin{subequations}
\begin{eqnarray}
a^0(\tau) &=& \frac{du^0(\tau)}{d\tau} = a\sinh(a\tau),\\
a^1(\tau) &=& \frac{du^1(\tau)}{d\tau} = a\cosh(a\tau).
\end{eqnarray}
\end{subequations}
The world line of a uniformly accelerating observer in flat Minkowski spacetime is indeed an acceleration 
curve: It is smooth, timelike and future directed, and the norm of the vector $a^\mu$ is
constant at every point of the curve:
\begin{equation}
\vert\vert a^\mu\vert\vert = \sqrt{a^\mu a_\mu} = \sqrt{-(a^0)^2 + (a^1)^2} 
= a\sqrt{-\sinh^2(a\tau) + \cosh^2(a\tau)} = a.
\end{equation}

       As another example of an accaleration curve we may consider the timelike 
curves in Schwarzschild spacetime, where all of the Schwarzschild coordinates 
$r$, $\theta$ and $\phi$ are constants such that $r>2M$. When we move along that 
curve, the only Schwarzschild coordinate having explicit dependence on the 
proper time $\tau$ is:
\begin{equation}
t(\tau) = (1 - \frac{2M}{r})^{-1/2}\,\tau,
\end{equation}
and hence the only non-zero component of $u^\mu(\tau)$ is:
\begin{equation}
u^t(\tau) = (1 - \frac{2M}{r})^{-1/2}.
\end{equation}
The only non-zero component of $a^\mu(\tau)$ is:
\begin{equation}
a^r(\tau) = \Gamma^r_{tt}(u^t(\tau))^2 = \frac{M}{r^2},
\end{equation}
and hence we find that the norm of the proper acceleration is:
\begin{equation}
\sqrt{a_r(\tau)a^r(\tau)} = (1 - \frac{2M}{r})^{-1/2}\,\frac{M}{r^2},
\end{equation}
which is constant at every point of the curve. So we indeed have an acceleration
curve.

\subsubsection{Properties of Acceleration Curves}

    It is an important property of acceleration curves that the vectors $a^\mu$ and $u^\mu$ are
orthogonal. In other words, we have
\begin{equation}
u^\mu(\tau)a_\mu(\tau) \equiv 0
\end{equation}
at every point of an acceleration curve. To show that this is indeed the case, let us pick up
an arbitrary acceleration curve, and an arbitrary point $P$ on that curve. At the point
$P$ of spacetime we may pick up an othonormal geodesic system of coordinates, where 
$\Gamma^\mu_{\alpha\beta}$ vanishes at $P$. Hence we have:
\begin{equation}
a^\mu(P) = \frac{du^\mu(P)}{d\tau}.
\end{equation}
Using the product rule we get:
\begin{equation}
u^\mu(P)a_\mu(P) = \frac{1}{2}\frac{d}{d\tau}(u^\mu(P)u_\mu(P)) = 0,
\end{equation}
because $u^\mu(\tau)u_\mu(\tau) \equiv -1$. Since $P$ is arbitrary, Eq.(A15) really implies 
Eq.(A.13).

\subsection{Acceleration Congruence}

  We shall now define the concept of {\it acceleration congruence} as a smooth congruence of 
acceleration curves parametrized by the proper time $\tau$ measured along the elements 
of the congruence such that:

(i) All those sets of points, where $\tau = constant$ along the elements of the congruence 
are smooth, orientable, simply connected, spacelike two-surfaces of spacetime.

(ii) The norm, or absolute value, of the proper acceleration vector of each element of 
the congruence is the same. 

(iii) For arbitrary, fixed $\tau$ the proper acceleration vector field
\begin{equation}
a^\mu := u^\alpha u^\mu_{;\alpha}
\end{equation}
of the congruence is parallel to a spacelike normal vector field of the spacelike
two-surface, where $\tau = constant$.

(iv) The spacelike two-surface, where $\tau = 0$, intersects orthogonally the elements
of the congruence.

\subsection{Acceleration Surface}

  After defining the concept of acceleration congruence we are able to define {\it acceleration
surface}, quite simply, as an equivalence class of those sets of points, where $\tau = constant$
along the elements of an acceleration congruence. By definition, the elements of these 
equivalence classes are smooth, orientable and simply connected, spacelike two-surfaces of
spacetime. If we pick up any two spacelike two-surfaces of spacetime with these properties, 
the surfaces are equivalent, i.e. they belong to the same equivalence class, if they are
$\tau=constant$ surfaces of the {\it same} acceleration congruence. In other words, acceleration
surfaces are labelled by the corresponding acceleration congruences. Physically, we may think 
an acceleration surface as a certain spacelöike two-surface propagating in spacetime. Because
of that the acceleration congruence determining a given acceleration surface constitutes the 
congruence of the world lines of the points of that surface.

   Our definition implies that acceleration surface has a spacelike unit normal vector 
field $n^\mu$ such that
\begin{equation}
a^\mu n_\mu \equiv constant := a
\end{equation}
at every point of an acceleration surface propagating in spacetime. Using Eqs.(A13) and (A17) we 
find that the unit vector fields $u^\mu$ and $n^\mu$ are orthogonal. In other words, we have:
\begin{equation}
u^\mu n_\mu \equiv 0.
\end{equation}
So we see that our mathematically precise definition of an acceleration surface reproduces 
Eq.(2.15) which was used, in Section II, as the starting point of our heuristic
definition.

\subsection{Examples of Acceleration Surfaces}

   It is very easy to give examples of acceleration surfaces. For instance, an equivalence 
class of the $t = constant$ slices of the timelike hypersurface, where $r = constant (>2M)$ 
in Schwarzschild spacetime is an acceleration surface. The timelike hypersurface, where
$r = constant (> 2M)$ consists of the points of the world lines of the observers at rest 
with respect to the Schwarzschild coordinates such that the radial coordinate $r$ of 
all these observers is the same. It is easy to see that these world lines constitute an 
acceleration congruence: The world lines are acceleration curves whose congruence is smooth, 
the sets of points, where $\tau = constant$ are spacelike two-spheres with radius $r$, 
which are smooth, orientable, simply connected spacelike two-surfaces of spacetime, and the
norm of the proper acceleration vector of each element of the congruence is the same, being
given by Eq.(A12). Moreover, the only non-zero component of the proper acceleration vector
field $a^\mu$ of the congruence is the component $a^r$ of Eq.(A11), and so the vector 
field $a^\mu$ of the congruence is parallel to the spacelike unit normal vector field
\begin{equation}
n^\mu := \delta^\mu_r(1 - \frac{2M}{r})^{1/2}
\end{equation}
of the spacelike two-sphere, where $r = constant$. Finally, if we parametrize the elements 
of the congruence by means of the proper time $\tau$ as in Eq.(A9) we find that the spacelike 
two-sphere, where $\tau = 0$, is orthogonal to the elements of the congruence. So we see that 
all points in the definition of the concept of acceleration surface are satisfied, and 
therefore the equivalence class of the $t = constant$ slices of the $r = constant$ hypersurface
is indeed an acceleration surface.

   As another example we may consider the set of points where $\tau = constant$ along the
world lines of uniformly accelerating observers in flat Minkowski spacetime such that these 
world lines are parametrized as in Eq.(A5). If we take that set of points, where $\tau = 0$ to 
be a plane parallel to the $yz$-plane such that the $y$- and the $z$-coordinates of the points
of that plane are the numbers of the interval $\lbrack 0,L\rbrack$, where $L>0$, it immediately
follows that the sets of points, where $\tau = constant$ are planes parallel to the 
$yz$-plane, and the equivalence class of those planes is an acceleration surface.

\subsection{Properties of Acceleration Surfaces}

    Acceleration surfaces have the following, very important property:

    {\bf Theorem A1}: {\it Acceleration surface intersects orthogonally the world lines 
of its points.}

   To prove this very important theorem let us fix the spacetime coordinates in such a way that
on the world lines of the points of an acceleration surface the time coordinate agrees with 
the proper time $\tau$ measured along those world lines, and those points on the two-surfaces 
$\tau = constant$ which belong to the same world line have the same spatial coordinates $x^1$
and $x^2$. In other words, the coordinates $x^1$ and $x^2$ provide a specific system of 
coordinates for the points of an acceleration surface propagating in spacetime. Let us denote
the tangent vectors of the corresponding coordinate curves on the acceleration surface by
$b^\mu_{(\alpha)}$, where $\alpha = 1, 2$. To prove our theorem we should show that
\begin{equation}
u_\mu b^\mu_{(\alpha)} \equiv 0
\end{equation}
for every instant of the proper time $\tau$ and at every point of our acceleration surface
for all $\alpha = 1, 2$. By definition, acceleration surface intersects the world lines of 
its points orthogonally, when $\tau = 0$, and therefore $u_\mu b^\mu_{(\alpha)} = 0$ for all $\alpha = 1, 2$, 
when $\tau = 0$. Hence it is sufficient to show that
\begin{equation}
\frac{d}{d\tau}(u_\mu b^\mu_{(\alpha)}) = 0
\end{equation}
for every $\tau$ and all $\alpha = 1, 2$. 

    The left hand side of Eq.(A21) may be written as:
\begin{equation}
\frac{d}{d\tau}(u_\mu b^\mu_{(\alpha)}) = u^\sigma(u_\mu b^\mu_{(\alpha)})_{;\sigma}
= u^\sigma u_{\mu;\sigma} b^\mu_{(\alpha)} + u^\sigma u_\mu b^\mu_{(\alpha);\sigma},
\end{equation}
wehere the first equality follows from the chain rule, and the second from the product rule of
covariant differentiation. The first term on the right hand side of Eq.(A22) may be written as:
\begin{equation}
u^\sigma u_{\mu;\sigma} b^\mu_{(\alpha)} = a_\mu b^\mu_{(\alpha)} = an_\mu b^\mu_{(\alpha)},
\end{equation}
where we have used Eqs.(A16) and (A17). Because $n^\mu$ is a spacelike unit normal of the 
acceleration surface, we have $n_\mu b^\mu_{(\alpha)} = 0$, and therefore the first term
on the right hand side of Eq.(A22) vanishes. Hence we are left with the second term only.

   To show that the second term on the right hand side of Eq.(A22) vanishes as well, we note 
first that in our system of spacetime coordinates $u^\mu$ is the tangent vector of the 
coordinate curve corresponding to the time coordinate. Because of that we have
\begin{equation}
u^\mu_{;\sigma} = \Gamma^\mu_{\sigma\beta} u^\beta,
\end{equation}
and since $b^\mu_{(1)}$ and $b^\mu_{(2)}$ are the tangent vectors of the coordinate curves
corresponding, respectively, to the coordinates $x^1$ and $x^2$, we have:
\begin{equation}
b^\mu_{(\alpha);\sigma} = \Gamma^\mu_{\sigma\beta} b^\beta_{(\alpha)},
\end{equation}
and we get:
\begin{subequations}
\begin{eqnarray}
u^\sigma b^\mu_{(\alpha);\sigma} &=& \Gamma^\mu_{\sigma\beta} u^\sigma b^\beta_{(\alpha)},\\
b^\sigma_{(\alpha)}u^\mu_{;\sigma} &=& \Gamma^\mu_{\sigma\beta} b^\sigma_{(\alpha)}u^\beta.
\end{eqnarray}
\end{subequations}
So we have:
\begin{equation}
u^\sigma b^\mu_{(\alpha);\sigma} = b^\sigma_{(\alpha)}u^\mu_{;\sigma},
\end{equation}
and we may write:
\begin{equation}
\frac{d}{d\tau}(u_\mu b^\mu_{(\alpha)}) = u_\mu u^\mu_{;\sigma} b^\sigma_{(\alpha)} 
= \frac{1}{2}(u_\mu u^\mu)_{;\sigma}b^\sigma_{(\alpha)} = 0,
\end{equation}
where we have used the product rule of covariant differentiation, and the fact that
$u_\mu u^\mu \equiv -1$. So we have shown that Eq.(A21), and therefore Eq.(A20) holds. In other words,
we have proved our theorem.

\subsection{Construction of Acceleration Surfaces}

  We shall use our theorem in Appendix B, where we consider the dynamical properties of
acceleration surfaces. Another important feature of the theorem lies in the fact that it tells
how one may {\it construct} an acceleration surface in arbitrary spacetime.

  When constructing an acceleration surface in arbitrary spacetime the first step is to 
pick up a smooth, orientable, spacelike two-surface of spacetime. The second step is to
pick up a future directed, smooth, timelike unit vector field $u^\mu$ orthogonal to the 
surface, and a smooth spacelike unit vector field $n^\mu$ orthogonal both to the surface 
and the vector field $u^\mu$. There is an infinite number of ways to do this choice, but once
after we have fixed the vector field $u^\mu$, the vector field $n^\mu$ is uniquely determined up to 
the sign. The different possible choices for the vectors $u^\mu$ and $n^\mu$ may be obtained
from each other by means of the Lorentz transformation. In other words, if the vector fields
$u^\mu$ and $n^\mu$ as well as the vector fields $u'^\mu$ and $n'^\mu$, are both allowed choices
for the timelike and spacelike unit normal vector fields of the surface, then between these vector
fields there is the relationship:
\begin{subequations}
\begin{eqnarray}
u'^\mu &=& u^\mu\cosh\phi + n^\mu\sinh\phi,\\
n'^\mu &=& u^\mu\sinh\phi + n^\mu\cosh\phi,
\end{eqnarray}
\end{subequations}
where $\phi$ is the boost angle.

  Since the vector field $u^\mu$ is orthogonal to our spacelike two-surface, our theorem implies
that $u^\mu$ may be taken to be a future directed unit tangent vector field of the congruence of 
the world lines of the points of an acceleration surface. As such the vector field $u^\mu$ tells 
in which direction we should move the points of our initial spacelike two-surface of spacetime.
More precisely, if the coordinates of a specific point on the spacelike two-surface, where the
proper time $\tau$ is constant are $x^\mu(\tau)$, the coordinates of that point on the 
two-surface, where the proper time is $\tau + d\tau$ are
\begin{equation}
x^\mu(\tau + d\tau) = x^\mu(\tau) + u^\mu(\tau)\,d\tau.
\end{equation}
On that two-surface the vector field $u^\mu$ is slightly different from what it was on our
initial surface. Since
\begin{equation}
a^\mu := u^\alpha u^\mu_{;\alpha} = an^\mu,
\end{equation}
we find that
\begin{equation}
u^\mu(\tau + d\tau) = u^\mu(\tau) + (an^\mu(\tau) 
- \Gamma^\mu_{\alpha\beta}(\tau)u^\alpha(\tau)u^\beta(\tau))\,d\tau.
\end{equation}
Our theorem ensures that for infinitesimal $d\tau$ the vector 
$u^\mu(\tau + d\tau)$ is orthogonal to the spacelike two-surface, where the proper time is 
$\tau + d\tau$. The vector $u^\mu(\tau + d\tau)$ tells in which direction we should move 
the points of that two-surface, and we may proceed as before. On the spacelike two-surface
with the proper time $\tau + d\tau$ the components of the spacelike unit vector field $n^\mu$
are
\begin{equation}
n^\mu(\tau + d\tau) = n^\mu(\tau) + (au^\mu(\tau) 
- \Gamma^\mu_{\alpha\beta}(\tau)n^\alpha(\tau)u^\beta(\tau))\,d\tau,
\end{equation}
which follows from the fact that $n^\mu$ obeys an equation
\begin{equation}
u^\alpha n^\mu_{;\alpha} = au^\mu.
\end{equation}
Moving the points of our initial spacelike two-surface with infinitesimal steps in spacetime
in the manner described above we get a sequence of smooth, orientable, simply connected 
spacelike two-surfaces of spacetime, whose equivalence class constitutes an acceleration 
surface.

\maketitle

\section{Dynamics of Acceleration Surfaces}

  Acceleration surfaces have {\it dynamics} in the sense that the geometrical properties of
an acceleration surface may change when its points propagate in curved spacetime. For instance,
the area of an acceleration surface may change. In this Appendix we consider in which way 
the changes in the area of an acceleration surface depend on the geometrical properties of the
underlying spacetime.

\subsection{Area of an Acceleration Surface}

  In general, the area of an acceleration surface in a specific instant of the proper time
$\tau$ measured along the world lines of its points is
\begin{equation}
A(\tau) = \int_{\mathcal{S}(\tau)}d\mathcal{A},
\end{equation}
where $d\mathcal{A}$ is the area element on the surface, and the integration has been performed
over the whole surface $\mathcal{S}(\tau)$ associated with the proper time $\tau$. 

   The first question is: How to calculate the area element $d\mathcal{A}$? To begin with, we
note that the points of the elements of an acceleration congruence, or the world lines of the 
points of an acceleration surface, constitute a three-dimensional timelike hypersurface of 
spacetime. To find a practical way of calculating $d\mathcal{A}$ let us fix the timelike
coordinate $x^0$ and the three spacelike coordinates $x^1$, $x^2$ and $x^3$ of spacetime
in such a way that the timelike coordinate $x^0$
coincides with the proper time $\tau$, the coordinate $x^3$ is constant on the hypersurface, 
and the coordinates $x^1$ and $x^2$ are the same for all those points which belong to the 
same world line, or acceleration curve. An example of this kind of a choice of coordinates 
is given by the Schwarzschild coordinates in Schwarzschild spacetime. We saw in the Appendix
A that the world lines of observers at rest with respect to the Schwarzschild coordinates, all 
having the same Schwarzschild coordinate $r(>2M)$, constitute an acceleration congruence. If
we take the coordinate $r$ to be the coordinate $x^3$, the coordinates $\theta$ and $\phi$, 
respectively, to be the coordinates $x^1$ and $x^2$, and replace the Schwarzschild time 
coordinate $t$ by the proper time $\tau$, we have constructed in Schwarzschild spacetime
a system of coordinates described above.

  According to the Theorem A1 of the Appendix A acceleration surface intersects orthogonally
the world lines of its points. This theorem implies that on the timelike hypersurface 
constituted by the points of the elements of an acceleration congruence our system of 
coordinates is time orthogonal. In other words, the line element on this hypersurface may be 
written  as:
\begin{equation}
d\tilde{s}^2 = -d\tau^2 + q_{mn}(\tau)\,dx^m\,dx^n,
\end{equation}
where $m, n = 1,2$, and $q_{mn}(\tau)$ is the metric induced on the spacelike two-surface,
where $\tau=constant$. We have denoted the line element by $d\tilde{s}^2$, intead of 
$ds^2$, because it tells, in our system of coordinates, the line element induced on a 
certain three-dimensional timelike hypersurface of spacetime, where the spacelike coordinate
$x^3=constant$. The four-dimensional spacetime metric $g_{\mu\nu}$, in turn, has the property
that
\begin{equation}
g_{a0} \equiv g^{a0} \equiv 0
\end{equation}
for every $a = 1, 2, 3$ on our timelike hypersurface.

   Using the metric tensor $q_{mn}(\tau)$ induced on the spacelike two-surface, where 
$\tau = constant$, one may write the area of an acceleration surface in a specific instant
of the proper time $\tau$ as:
\begin{equation}
A(\tau) = \int_{\mathcal{S}(\tau)}\sqrt{q(\tau)}\,d^2x,
\end{equation}
where $q(\tau)$ is the determinant of the metric. Since the coordinates $x^1$ and $x^2$ are 
constants along the world lines of the points of the acceleration surface, the domain of 
integration for the coordinates $x^1$ and $x^2$ remains the same for all $\tau$. Because
of that we may write the first proper time derivative of the area $A$ of an acceleration 
surface as:
\begin{equation}
\frac{dA(\tau)}{d\tau} = \int_{\mathcal{S}}(\frac{d}{d\tau}\sqrt{q(\tau)})\,d^2x,
\end{equation}
or:
\begin{equation}
\frac{dA}{d\tau} = \frac{1}{2}\int_{\mathcal{S}} q^{mn}\dot{q}_{mn}\sqrt{q}\,d^2x,
\end{equation}
where the dot denotes the proper time derivative, and $q^{mn}$ the inverse of $q_{mn}$. To 
simplify the notation we have dropped off the references to the proper time $\tau$. 

\subsection{The First Proper Time Derivative of the Area} 

  A useful expression for the first proper time derivative of the area of an acceleration surface
is:
\begin{equation}
\frac{dA}{d\tau} = \int_{\mathcal{S}} u^{\mu;\nu}\gamma_{\mu\nu}\,d\mathcal{A},
\end{equation}
where we have defined the tensor $\gamma_{\mu\nu}$ as:
\begin{equation}
\gamma_{\mu\nu} := g_{\mu\nu} + u_\mu u_\nu - n_\mu n_\nu.
\end{equation}
As in Appendix A, $u^\mu$ is the future directed unit tangent vector field of the congruence
of the world lines of the points of the acceleration surface, and $n^\mu$ is the spacelike unit 
normal, orthogonal to $u^\mu$, of the surface. The tensor $\gamma_{\mu\nu}$ projects vectors on 
the acceleration surface. If we pick up an arbitrary vector $A^\mu$ of spacetime, then 
$\gamma^\mu_{\,\,\nu}A^\nu$ is the projection of that vector on the acceleration surface.

  Eq.(B7) is a tensorial equality, and therefore it holds in any system of coordinates, provided
that we manage to find {\it one} system of coordinates, where it holds. Hence it is sufficient 
to prove Eq.(B7) in the system of coordinates we defined above. We denote by $b^\mu_m$ 
$(m=1, 2)$ the coordinate basis vector fields corresponding to the coordinates $x^1$ and $x^2$
on the acceleration surface. As such they have the following properties:
\begin{subequations}
\begin{eqnarray}
b^\mu_m b_{n\mu} &=& q_{mn},\\
b^m_\mu b_{m\nu} &=& \gamma_{\mu\nu}.
\end{eqnarray}
\end{subequations}
The Greek indices $\mu$ and $\nu$ are pushed up and down by the spacetime metric $g_{\mu\nu}$,
and the Latin indices $m$ and $n$ by means of the two-metric $q_{mn}$ induced on the 
acceleration surface. The components of the coordinate basis vector fields $b^\mu_m$ are:
\begin{equation}
b^\mu_m = \delta^\mu_m
\end{equation}
for every $\mu = 0, 1, 2, 3$ and $m = 1, 2$. The only non-zero component of $u^\mu$, in turn,
is:
\begin{equation}
u^0 = -u_0 = 1.
\end{equation}
The integrand on the right hand side of Eq.(B7) may be written as:
\begin{equation}
u_{\mu;\nu}\gamma^{\mu\nu} = u_{\mu;\nu}b^\mu_m b^{m\nu} = \Gamma^0_{mn}q^{mn} 
= \frac{1}{2}q^{mn}\dot{q}_{mn}.
\end{equation}
The first equality follows from Eq.(B9b), the second from Eqs.(B10) and (B11), and the third
from Eq.(B3), together with the facts that in our system of coordinates 
$g_{00} = g^{00} = -1$, and the components $g_{mn}$ of the spacetime metric agree with the
components $q_{mn}$ of the metric iduced on the acceleration surface. Hence we find that Eq.(B7)
implies:
\begin{equation}
\frac{dA}{d\tau} = \frac{1}{2}\int_{\mathcal{S}} q^{mn}\dot{q}_{mn}\sqrt{q}\,d^2x,
\end{equation}
which is exactly Eq.(B6). In other words, we have proved that the first proper time derivative
of the area of an acceleration surface, when it propagates in curved spacetime, is indeed given 
by Eq.(B7).

   There are different ways of writing the right hand side of Eq.(B7). An expression which will 
become useful in a moment is:
\begin{equation}
\frac{dA}{d\tau} = \int_{\mathcal{S}} u^\mu_{;\nu}E^I_\mu E^\nu_I\,d\mathcal{A},
\end{equation}
where the two spacelike vector fields $E^\mu_I$ $(I = 1, 2)$ are orthonormal tangent vector
fields on the acceleration surface. In other words, the fields $E^\mu_I$ have the properties:
\begin{subequations}
\begin{eqnarray}
E^\mu_I E_{J\mu} &=& \delta_{IJ},\\
E^I_\mu E_{I\nu} &=& \gamma_{\mu\nu},
\end{eqnarray}
\end{subequations}
for all $I, J = 1, 2$, and $\mu, \nu = 0, 1, 2, 3$. Again, the Greek indices $\mu$ and $\nu$ are
pushed up and down by means of the spacetime metric $g_{\mu\nu}$, and the upper case Latin 
indices $I$ and $J$ by $\delta_{IJ}$. Theorem A1 implies that the vector fields $E^\mu_I$,
together with the vector fields $u^\mu$ and $n^\mu$, constitute a base of a four-dimensional
othonormal system of coordinates at each point of the acceleration surface.

   The first proper time derivative of the area of an acceleration surface depends, in addition
to the underlying spacetime geometry, on the {\it initial conditions} posed for the future 
directed tangent vector field $u^\mu$ of the corresponding acceleration congruence. For 
instance, we may pose for the field $u^\mu$ an initial condition:
\begin{equation}
u^\mu_{;\alpha}E^\alpha_I = 0
\end{equation}
for all $I = 1, 2$ at every point of an acceleration surface, when $\tau = 0$. Eq.(B16) states
that the vectors $u^\mu$ associated with the different points of an acceleration  surface are 
parallel to each other, when $\tau = 0$, and it has an important consequence: Using Eq.(B14)
we find that when Eq.(B16) holds, the first proper time derivative of the area of an 
acceleration surface vanishes, when $\tau = 0$, i.e. 
\begin{equation}
\frac{dA}{d\tau}\vert_{\tau=0} = 0.
\end{equation}
In the time orthogonal system of coordinates used on the acceleration congruence the initial 
condition (B16) implies that:
\begin{equation}
\frac{d\sqrt{q}}{d\tau}\vert_{\tau=0} = 0,
\end{equation}
which follows from the fact that in our specific system of coordinates the left hand side of 
Eq.(B18) is exactly the integrand of Eq.(B14) which, in turn, vanishes identically when $\tau=0$ 
by means of Eq.(B16).

\subsection{The Second Proper Time Derivative of the Area}

   When Eq.(B16) holds, one finds that the {\it second} proper time derivative of the
area of an acceleration surface may be written, when $\tau=0$, as:
\begin{equation}
\frac{d^2A}{d\tau^2}\vert_{\tau=0} 
= \int_{\mathcal{S}} u^\alpha u^\mu_{;\nu;\alpha}E^I_\mu E^\nu_I\,d\mathcal{A}.
\end{equation}
Again, we have a tensorial equation which is proved in a general system of coordinates, 
provided that just one system of coordinates, where that equation holds, is found. In our
time orthogonal system of coordinates we may write, because the domain of integration remains
the same for all $\tau$:
\begin{equation}
\frac{d^2A}{d\tau^2}\vert_{\tau=0} 
= \int_{\mathcal{S}}\frac{d}{d\tau}(u^{\mu;\nu}\gamma_{\mu\nu}\sqrt{q})\vert_{\tau=0}\,d^2x,
\end{equation}
where we have used Eq.(B7). Using Eq.(B18) we find:
\begin{equation}
\frac{d}{d\tau}(u^{\mu;\nu}\gamma_{\mu\nu}\sqrt{q})\vert_{\tau=0} 
= \frac{d}{d\tau}(u^{\mu;\nu}\gamma_{\mu\nu})\vert_{\tau=0}\sqrt{q},
\end{equation}
and the chain rule implies:
\begin{equation}
\frac{d}{d\tau}(u^{\mu;\nu}\gamma_{\mu\nu}) = u^\alpha(u^{\mu;\nu}\gamma_{\mu\nu})_{,\alpha}
= u^\alpha(u^{\mu;\nu}\gamma_{\mu\nu})_{;\alpha}.
\end{equation}
We have been allowed to replace the ordinary partial derivatives by the covariant ones, because
the function inside the brackets is a scalar. According to the product rule we have:
\begin{equation}
u^\alpha(u^{\mu;\nu}\gamma_{\mu\nu})_{;\alpha} = u^\alpha u^{\mu;\nu}_{\,\,\,\,\,\,;\alpha}\gamma_{\mu\nu}
+ u^\alpha u^{\mu;\nu}\gamma_{\mu\nu;\alpha},
\end{equation}
and it follows from Eq.(B8), the definition of $\gamma_{\mu\nu}$, that:
\begin{equation}
u^\alpha\gamma_{\mu\nu;\alpha} = (u^\alpha u_{\mu;\alpha})u_\nu + u_\mu(u^\alpha u_{\nu;\alpha}) 
- (u^\alpha n_{\mu;\alpha})n_\nu - n_\mu(u^\alpha n_{\nu;\alpha}).
\end{equation}
However, using Eqs.(A16), (A17) and (A34) we find:
\begin{equation}
u^\alpha\gamma_{\mu\nu;\alpha} = a(n_\mu u_\nu + u_\mu n_\nu - u_\mu n_\nu - n_\mu u_\nu) = 0,
\end{equation}
and so the second term on the right hand side of Eq.(B23) vanishes. As a result we get, using Eqs.(B20)
and (B21):
\begin{equation}
\frac{d^2A}{d\tau^2}\vert_{\tau=0} = \int_{\mathcal{S}} u^\alpha u^{\mu;\nu}_{\,\,\,\,\,\,;\alpha}
\gamma_{\mu\nu}\sqrt{q}\,d^2x,
\end{equation}
which readily implies Eq.(B19) by means of Eq.(B15b) and the fact that $d\mathcal{A} = \sqrt{q}\,d^2x$
in our system of coordinates.

   It is possible to write Eq.(B19) in a form which provides a direct connection with the underlying 
spacetime geometry. Using the trivial identity
\begin{equation}
u^\alpha u_{\mu;\nu;\alpha} = u^\alpha u_{\mu;\alpha;\nu} - u^\alpha(u_{\mu;\alpha;\nu} - u_{\mu;\nu;\alpha}),
\end{equation}
together with the basic properties of the Riemann tensor and the product rule of covariant differentiation we get:
\begin{equation}
u^\alpha u_{\mu;\nu;\alpha} = (u^\alpha u_{\mu;\alpha})_{;\nu} - u^\alpha_{;\nu}u_{\mu;\alpha} 
- u^\alpha R^\beta_{\,\,\mu\alpha\nu}u_\beta.
\end{equation}
Applying Eq.(A30), the definition of the proper acceleration vector field $a^\mu$ of our acceleration
congruence, and the symmetry properties of the Riemann tensor we find:
\begin{equation}
u^\alpha u^\mu_{\,\,;\nu;\alpha} = a^\mu_{\,\,;\nu} - u^\alpha_{;\nu} u^\mu_{;\alpha} 
+ R^\mu_{\,\,\alpha\beta\nu} u^\alpha u^\beta.
\end{equation}
Now, we have
\begin{equation}
u^\alpha_{;\nu}u^\mu_{;\alpha}E^I_\mu E_I^\nu = 0,
\end{equation}
provided that Eq.(B.16) holds. Moreover,
\begin{equation}
R^\mu_{\,\,\alpha\beta\nu}E^I_\mu E^\nu_I = R_{\alpha\beta}u^\alpha u^\beta 
+ R_{\mu\alpha\beta\nu}u^\mu u^\alpha u^\beta u^\nu - R_{\mu\alpha\beta\nu}n^\mu n^\nu u^\alpha u^\beta.
\end{equation}
The second term on the right hand side of Eq.(B31) vanishes because of the symmetry properties of the Riemann tensor.
So we get:
\begin{equation}
\frac{d^2A}{d\tau^2}\vert_{\tau=0} = \int_{\mathcal{S}}(a^\mu_{;\nu}E^I_\mu E^\nu_I + R_{\mu\nu}u^\mu u^\nu 
- R_{\alpha\mu\nu\beta}n^\alpha n^\beta u^\mu u^\nu)\,d\mathcal{A}
\end{equation}
which, by means of Eq.(A30), takes the form:
\begin{equation}
\frac{d^2A}{d\tau^2}\vert_{\tau=0} = \int_{\mathcal{S}}(ak_n + R_{\mu\nu}u^\mu u^\nu 
- R_{\alpha\mu\nu\beta}n^\alpha n^\beta u^\mu u^\nu)\,d\mathcal{A},
\end{equation}
where
\begin{equation}
k_n := n^\mu_{;\nu}E^I_\mu E^\nu_I
\end{equation}
is the trace of the exterior curvature tensor projected on the acceleration surface in a direction determined by the spacelike
unit normal $n^\mu$ of the surface.

   As one may observe, we have been able to write the second proper time derivative of the area of an acceleration
surface entirely in terms of the Ricci and the Riemann tensors of spacetime, together with the fields $u^\mu$ and 
$n^\mu$, except that we still have a term which depends on the intrinsic properties of the surface, i.e. on
its exterior curvature. To make even that term to vanish, we may pose yet another initial condition for the 
congruence of the world lines of the points of the acceleration surface, when $\tau=0$:
\begin{equation}
n^\mu_{;\nu}E^\nu_I = 0.
\end{equation}
This initial condition means that the exterior curvature of the acceleration surface vanishes, when $\tau=0$.
As such Eq.(B35) may be viewed as a definition of a "plane-like" acceleration surface. Indeed, a plane accelerating
in flat Minkowski spacetime with a constant proper acceleration to the direction of its spacelike normal satisfies
both of the initial conditions (B16) and (B35). In contrast, the $t=constant$ slices of the timelike hypersurface,
where $r=constant$ $(>2M)$ in Schwarzschild spacetime do {\it not} satisfy the initial condition (B35), although they 
do satisfy the initial condition (B16). The reason for that is that the $t=constant$ slices in question
are two-spheres, and the exterior curvature of a two-sphere embedded in Schwarzschild spacetime is non-zero.

   When both of the initial conditions (B16) and (B35) are satisfied, Eq.(B33) implies:
\begin{equation}
\frac{d^2A}{d\tau^2}\vert_{\tau=0} = \int_{\mathcal{S}}(R_{\mu\nu}u^\mu u^\nu 
- R_{\alpha\mu\nu\beta}n^\alpha n^\beta u^\mu u^\nu)\,d\mathcal{A}.
\end{equation}
This is the final result of this Appendix. It tells in which way the second proper time derivative of the area of 
an acceleration surface depends on the geometrical properties of the underlying spacetime, when the initial conditions
posed for the surface are chosen in such a way that all changes in the area are caused merely by the spacetime 
curvature, instead of being consequences of some specific choice of the initial conditions.

\end{document}